\documentclass[review,authoryear]{elsarticle}

\usepackage[latin1]{inputenc}

\usepackage{subfigure}
\usepackage{amssymb,amsmath,amsthm,amsfonts} 
\usepackage{graphicx,rotating} 
\usepackage{verbatim} 
\usepackage{enumerate,graphics}
\usepackage[dvips]{epsfig}
\usepackage{multirow}
\usepackage{url} 
\usepackage{hyperref} 
\usepackage{graphicx}
\usepackage{bm}
\usepackage{color}
\usepackage{placeins}

\usepackage{fullpage}
\allowdisplaybreaks[4] 

\newcommand{\E}{\operatorname{E}} 
\newcommand{\Var}{\operatorname{Var}} 
\newcommand{\R}{\mathbb R} 

\journal{Computers \& Industrial Engineering} 

\begin{document}

\begin{frontmatter} 

\title{Beta regression control chart for monitoring fractions and proportions
} 
 \author[bayer]{F\'abio~Mariano~Bayer\corref{cor1}}
 \ead{bayer@ufsm.br}
 \address[bayer]{Departamento de Estat\'istica and LACESM, Universidade Federal de Santa Maria, Brazil}    
 \cortext[cor1]{Corresponding author} 
     
 \author[tondolo]{Catia~Michele~Tondolo}
 \ead{catiatondolo@gmail.com}
 \address[tondolo]{IME, Universidade de S{\~a}o Paulo, Brazil}
 
 \author[muller]{Fernanda~Maria~M\"{u}ller}
 \ead{nandamuller90@gmail.com}
  \address[muller]{PPGA, Universidade Federal do Rio Grande do Sul, Brazil}

\begin{abstract} Regression control charts are usually used to monitor variables of interest that are related to control variables. However, for fraction and/or proportion data, the use of standard regression control charts may not be adequate, since the linear regression model assumes the normality of the interest variable. To work around this problem, we propose the beta regression control chart (BRCC). The BRCC is useful for monitoring fraction, rate and/or proportion data sets when they are related to control variables. The proposed control chart assumes that the mean and dispersion parameters of beta distributed variables are related to the exogenous variables, being modeled using regression structures. The BRCC is numerically assessed through an extensive Monte Carlo simulation study, showing good performance in terms of average run length (ARL). Two applications to real data are presented, evidencing the practical applicability of the proposed method. 

\end{abstract} 


\end{frontmatter} 

\section{Introduction}

The most usual application to monitor fraction or proportion data type consists of control charts for attributes of $p$ and $np$ types \citep{Oakland2007, Montgomery:2009}. These charts assume that the distribution of the nonconforming fraction follows a binomial distribution. The control limits for the $p$ chart are determined by $\text{CL}= \tilde{p} \pm w \sqrt{\frac{\tilde{p}(1-\tilde{p})}{n}}$, where $w$ is a constant that defines the width of the control limits corresponding to a control region (or, the number of standard deviations from the mean process), and $\tilde{p}$ is the mean. When the sample size $n$ is large, the binomial distribution will be approximately symmetric around the mean and the control limits can be calculated using an approximation to the normal distribution  \citep{Wang20094210, SantAnnaaandCaten:2012}.  The $np$ chart is very similar to the $p$ chart, 
one being a simply scaled version of the other. 

In situations where the number of defective items is small, $p$ and $np$ control charts are inaccurate in monitoring the process \citep{Wang20094210}. Another disadvantage is that the lower and upper limits can assume values outside the $(0,1)$ range, which has no physical meaning. In this case, the potential of the chart to detect process improvements may be compromised \citep{Bersimisetal:2014}.
Recent advances in control charts for monitoring fraction or proportion data type are found in \cite{chiu2015monitoring}, \cite{joekes2015extending}, \cite{um2016simultaneous}, and \cite{Raubenheimeretal:2016}. Despite advances, another difficulty arises when
 productive processes can be described by several characteristics, rather than a single quality characteristic \citep{CapizziandMasarotto:2011, Hawkins:1991}. In these cases, the process variables need to be monitored simultaneously \citep{CapizziandMasarotto:2011}. One way suggested in literature for monitoring a process of this type is to use regression control charts (RCC) \citep{Mandel:1969, Ryan:1989, HAWORTH:1996}.

Regarding the monitoring of the percentage of non-conforming items, the modeling using linear regression  is not always adequate, since the model requires that the dependent variable follows a normal distribution. The use of linear models in rate or proportion data can also generate predicted values outside the range $(0,1)$ \citep{Kieschnick2003a}. In addition, variables of the fraction and proportion type usually present asymmetry, which can compromise the inferences assuming erroneously data normality\citep{FerrariandCribari:2004}. For asymmetric data, 
an increase in the false alarm rate in linear regression control charts is also observed, due to the shape discrepancy of the data distribution using the normal distribution \citep{SantAnnaaandCaten:2012}.

As an alternative, to model processes of the fraction or proportion type, \cite{SantAnnaaandCaten:2012} proposed the beta control chart (BCC), determining the control limits through the quantile of the beta distribution. The beta distribution is flexible to model proportions and its density can take different shapes \citep{FerrariandCribari:2004}. However, BCC still does not allow 
for consideration of quality characteristics that influence the beta distributed variable of interest, since it considers the mean and dispersion parameters constant throughout the observations. To fill this gap we propose the beta regression control chart (BRCC), which allows to control the mean and the dispersion of variables of the fraction and proportion type in the presence of control variables.

The proposed BRCC considers the beta regression model with varying dispersion \citep{Simasetal:2010, Cribari-NetoandSouza:2012, BayerandCribari:2014}, where it is assumed that the dispersion parameter ($\sigma$) of the beta distribution is non constant throughout the observations, being modeled through a regression structure, in the same way as the mean ($\mu$). The BRCC generalizes the BCC presented by \cite{SantAnnaaandCaten:2012}, since it considers that $\mu$ and $\sigma$ are not constant throughout the observations. Thus, the proposed method allows that beta distributed quality characteristics are  related to control variables (covariates), similar to the linear regression control charts.

The monitoring of the mean and the dispersion of processes is relevant for providing robustness against the modeling of errors and unforeseen behaviors \citep{CapizziandMasarotto:2011}, being of interest to the statistical process control (SPC) to jointly control the mean and the dispersion of processes \citep{Teyarachakuletal:2007}. The increase of variability of a process may imply an increase of defective units, whereas a reduction of the dispersion may indicate an increase in process capacity, since more units will be close to the correct specifications \citep{raey, Huwangetal:2010, Zhangetal2015}. The correct modeling of the dispersion directly implies the determination of the control limits of the chart, in such a way that the incorrect specification of the dispersion can generate a high number of false alarms or loss of detection power of special causes. In these cases, it is not possible to analyze whether the process mean is under control when the dispersion is not under statistical control \citep{Huwangetal:2010}. Moreover, modeling the dispersion is necessary in regression models, in order to obtain accurate inferences about the structure parameters of the mean regression \citep{SmythandVerbyla:1999}.

To assess the performance of the proposed BRCC, a comparative study was carried out among: (i) the BRCC with varying dispersion, (ii) the BRCC with constant dispersion (BRCC$_C$), and (iii) the standard RCC \citep{Mandel:1969}. The BRCC$_C$ is a particular case of the proposed control chart, where only the mean of the distributed beta process is modeled and the dispersion parameter is considered constant throughout the observations. The comparison was performed numerically by means of Monte Carlo simulations, assessing the average run lengths (ARL) under control (ARL$_0$) and out-of-control (ARL$_1$). The numerical results show that the proposed control chart shows good performance in the detection of out-of-control points.

The paper is structured as follows. In Section~\ref{S:modelo}, some aspects related to beta control chart are discussed and the proposed beta regression control chart is presented. In Section~\ref{S:resultados}, the procedures of the sensitivity analysis of control charts are described, as well as the numerical results of the study of Monte Carlo simulation. In Section~\ref{Exemplo}, two applications are performed to actual data related to a tire manufacturing process and to the relative air humidity data of the city of Bras{\'i}lia, Brazil. Finally, Section~\ref{S:conclusion} describes the main conclusions of the paper.

\section{Beta regression control chart}\label{S:modelo} 

This section is divided into two subsections. Subsection~\ref{ss:bcc} presents a BCC review \citep{SantAnnaaandCaten:2012}, which is useful to statistically control variables of the fraction or proportion type that are independent and without the presence of control variables. Subsequently, Subsection~\ref{ss:proposed} presents the proposed BRCC. In addition to the definition of control limits, some aspects of inference and model selection in beta regression are also explored. 

\subsection{Review of beta control chart}\label{ss:bcc} 

Beta distribution is used to model continuous variables limited in the $(0,1)$ range, such as rates, fractions, and proportions. However, it is also useful when the interest variable $y$ is restricted to continuous interval $(a,b)$, where $a$ and $b$ are known scalars, and $a<b$. In these cases, without loss of generality, $(y-a)/(b-a)$ can be modeled instead of modeling $y$ directly. Being $y$ a random variable with beta distribution, its probability density function (PDF) is given by \citep{johnson1995continuous}:  
\begin{equation}\label{eq:fdpbeta1}
\pi(y;\theta_1,\theta_2) = {\Gamma(\theta_1+\theta_2)\over{\Gamma(\theta_1)\Gamma(\theta_2)}}y^{(\theta_1-1)}(1-y)^{(\theta_2-1)},  0<y<1,
\end{equation}
where $\theta_1 > 0$ and $\theta_2 >0$ are the parameters that index the distribution and $\Gamma(\cdot)$ is the gamma function. The mean and variance of $y$ are given, respectively, by: 
\begin{align}
\E(y)&={\theta_1\over{\theta_1+\theta_2}},  \\
\Var(y)&=\frac{\theta_1\theta_2}{(\theta_1+\theta_2)^2(\theta_1+\theta_2+1)}.
\end{align} The cumulative distribution function (CDF) of $y$ is \citep{Gupta2004}: 
\begin{equation}\label{eq:cdf}
F(y; \theta_1, \theta_2)= \int\limits_0^y \pi(u; \theta_1, \theta_2)du =\frac{B(y;\theta_1,\theta_2)}{B(\theta_1,\theta_2)},
\end{equation}
where $B(\theta_1,\theta_2)$ is the beta function and $B(y;\theta_1,\theta_2)$ is the incomplete beta function. The quantile function is given by $\psi(\alpha;\theta_1, \theta_2)=F^{-1}(\alpha; \theta_1, \theta_2)$. More details about beta distribution can be seen in \cite{Gupta2004}. 

Beta distribution is very versatile, having a wide variety of applications \citep{johnson1995continuous, Bury1999, Gupta2004}. In particular, \cite{SantAnnaaandCaten:2012} assume that variables of the fraction and proportion type have beta distribution, proposing the beta control chart. The BCC naturally accommodates the asymmetry of fraction and proportion data type, and the control limits become restricted to the $(0,1)$ range. These characteristics are advantageous compared to the usual charts that assume normal approximation, such as the $p$ and $np$ charts, which are popular for monitoring non-conforming items \citep{Shewhart1931}. 

The lower control (LCL) and upper control (UCL) limits of the beta control chart are defined by \citep{SantAnnaaandCaten:2012}:  
\begin{align}
\mbox{LCL}&=\bar{p} - w_1\sqrt{s^2(\bar{p})}, \label{e:licc}\\
\mbox{UCL}&=\bar{p} + w_2\sqrt{s^2(\bar{p})}, \label{e:lscc} 
\end{align}
where $\bar{p}$ and $s^2(\bar{p})$ are the mean and the variance of the fraction variable, $w_1$ and $w_2$ are tabulated constants, which define the width of the control limits. The values of $w_1$ and $w_2$ can be given by: 
\begin{align}
w_1&=\frac{\bar{p} - \psi([\alpha/2],\theta_1,\theta_2)}{\sqrt{s^2(\bar{p})}},  \label{e:w1}\\
w_2&=\frac{ \psi([1 - \alpha/2],\theta_1,\theta_2)-\bar{p} }{\sqrt[]{s^2(\bar{p})}}, \label{e:w2}
\end{align} 
for a control region $1-\alpha$ associated with a fixed average run length under control (ARL$_0=1/\alpha$). However, by replacing \eqref{e:w1} in \eqref{e:licc} and \eqref{e:w2} in \eqref{e:lscc}, these control limits can be directly rewritten as:  
\begin{align}
\mbox{LCL}&=\psi([\alpha/2],\theta_1,\theta_2) , \\
\mbox{UCL}&=\psi([1-\alpha/2],\theta_1,\theta_2),
\end{align}
where, in practice, $\theta_1$ and $\theta_2$ can be replaced by their respective estimates. In this paper, we will consider maximum likelihood estimators (MLE) for $\theta_1$ and $\theta_2$. MLE present, under usual conditions of regularity, good asymptotic properties \citep{Pawitan2001}, and they are the standard procedure  for parameters estimation in beta regression model \citep{FerrariandCribari:2004,Cribari-NetoandSouza:2012}.

However, the beta control chart does not consider situations where the practitioner is required to impose a regression structure for the interest variable. Our interest lies in situations where the mean and dispersion of the variable can be modeled as functions of a set of control variables and unknown parameters. 

\subsection{Proposed control chart}\label{ss:proposed} 

As in \cite{Cribari-NetoandSouza:2012} and \cite{BayerandCribari:2014, Bayer2015}, we consider a reparametrization of the beta density, where $\mu=\frac{\theta_1}{\theta_1+\theta_2}$ and $\sigma^2=\frac{1}{1+\theta_1+\theta_2}$, that is, $\theta_1= \mu\left(\frac{1-\sigma^2}{\sigma^2} \right)$ and $\theta_2=(1-\mu) \left(\frac{1-\sigma^2}{\sigma^2} \right)$. Assuming that $y$ follows a beta distribution with mean $\mu$ and dispersion parameter $\sigma$, the density of $y \sim Beta(\mu,\sigma)$ can be written as follows:  
\begin{align}\label{e:a}
f(y;\mu,\sigma)= \frac{\Gamma \left(\frac{1-\sigma^2}{\sigma^2}\right)}{\Gamma\left(\mu\left(\frac{1-\sigma^2}{\sigma^2}\right)\right)\Gamma\left(\left(1-\mu)(\frac{1-\sigma^2}{\sigma^2}\right)\right)}y^{\mu \left(\frac{1-\sigma^2}{\sigma^2}\right)-1}(1-y)^{(1-\mu) \left(\frac{1-\sigma^2}{\sigma^2}\right)-1},
\end{align}
where $0 < y < 1$, $0< \mu <1$ and $0< \sigma <1$. The mean and variance of $y$ are given, respectively, by:  
\begin{align}
\E(y)&=\mu,\\
\Var(y)&=\mu(1-\mu)\sigma^2. 
\end{align}
The cumulative distribution function of $y$ with this parametrization is given by: 
\begin{align}
\mathcal{F}(y; \mu_t, \sigma_t)= \int\limits_0^y f(u; \mu_t, \sigma_t)du= F\left(y;  \mu\left(\frac{1-\sigma^2}{\sigma^2} \right), (1-\mu) \left(\frac{1-\sigma^2}{\sigma^2} \right) \right).
\end{align}
This way, the quantile function is given by $Q(\alpha;\mu_t, \sigma_t)=\mathcal{F}^{-1}(\alpha;\mu_t,\sigma_t)$.
 
\begin{figure}[h]
\begin{center} 
\subfigure[
Different values of $\mu$ (indicated in the graph) with $\sigma=0.1$]{ 
\label{f:a1} \includegraphics [width=0.45\textwidth]{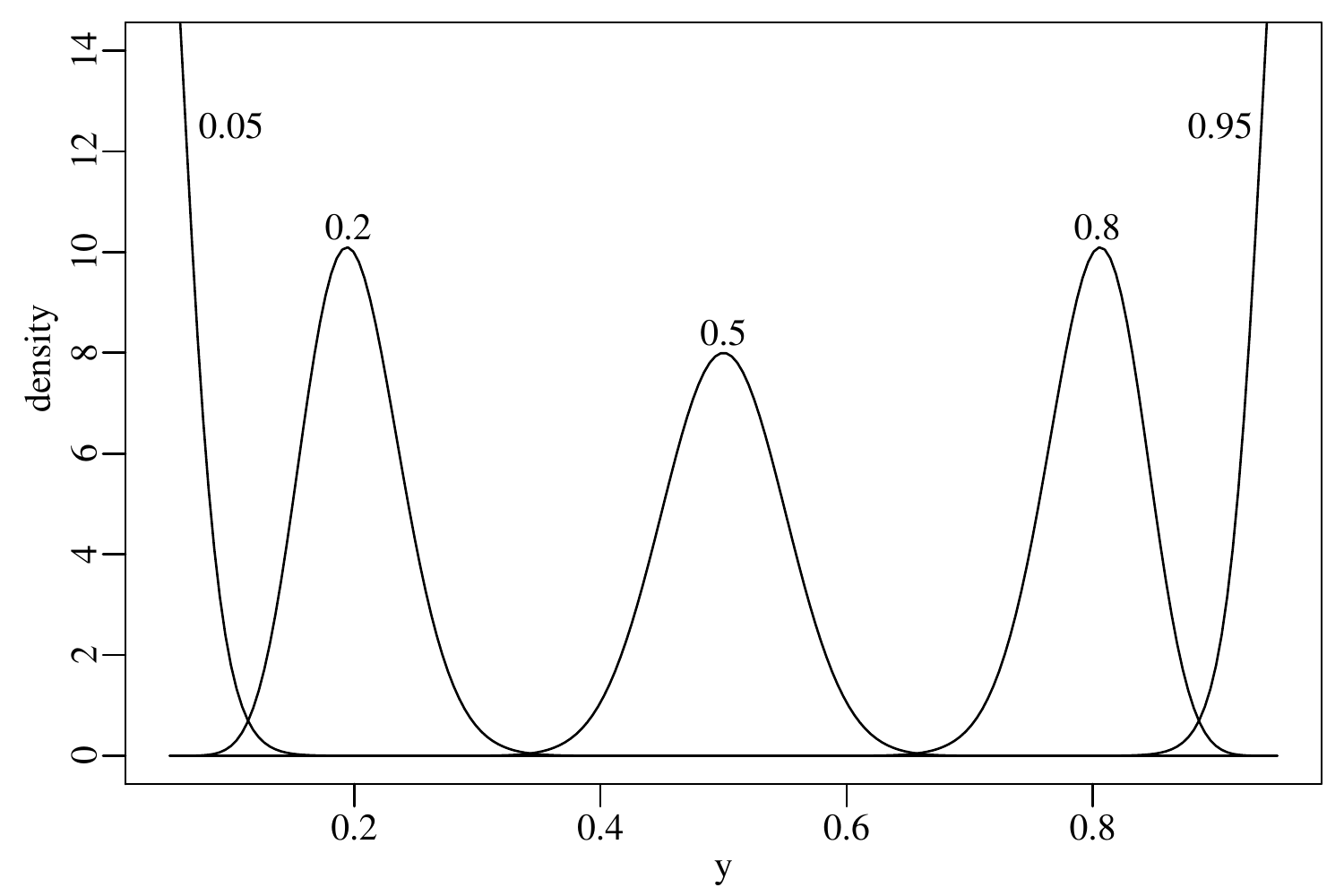}}
 \qquad 
\subfigure[Different values of $\sigma$ (indicated in the graph) with $\mu=0.4$]{ 
\label{f:b1} \includegraphics [width=0.45\textwidth]{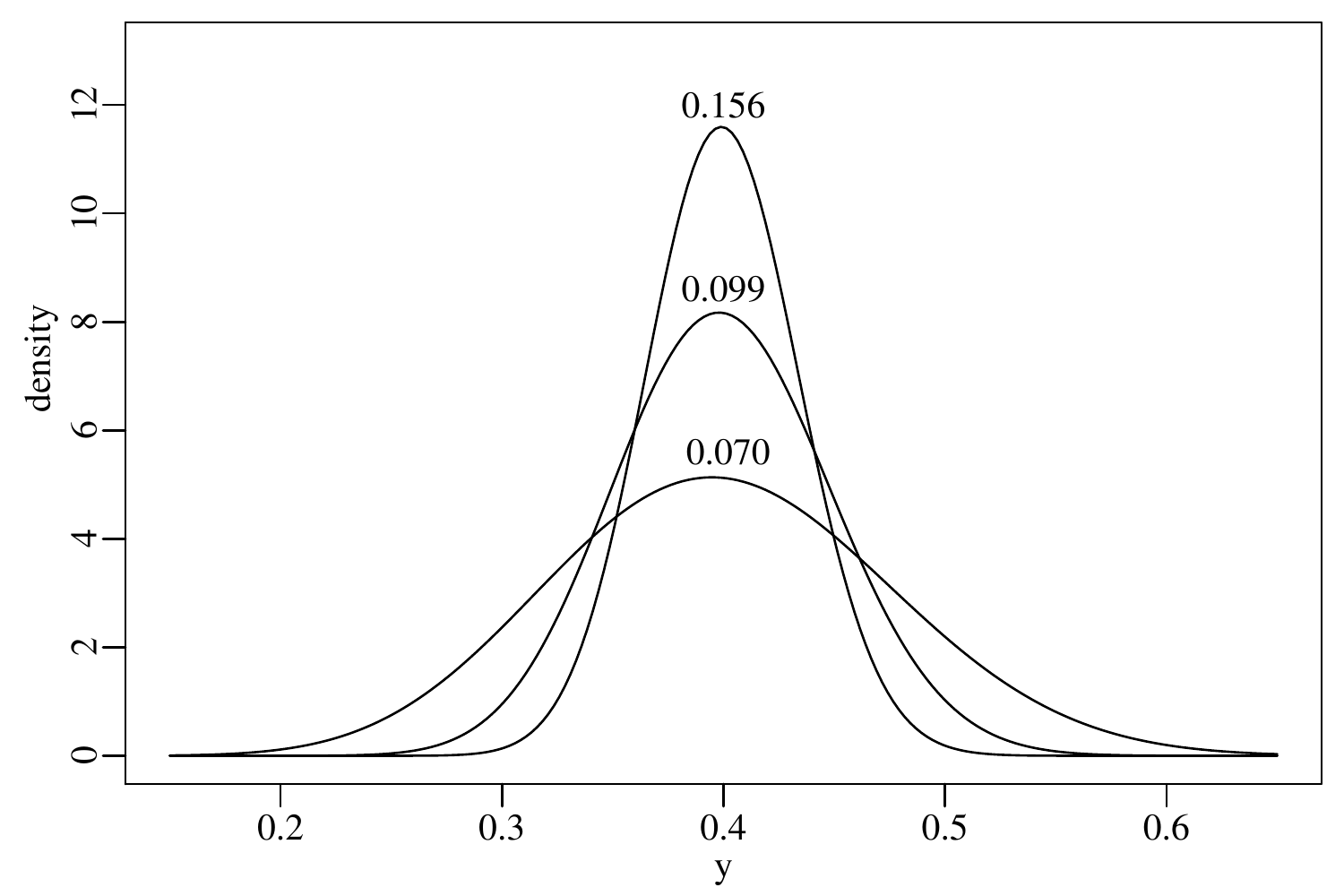}} 
\caption{Beta densities for different values of parameters $\mu$ and $\sigma$. 
}
\label{grafdens}
\end{center}
\end{figure}

The beta distribution is appropriate for rate and proportion type of data because of the variety of shapes (symmetric or asymmetric) that the density can take \citep{Kieschnick2003a}. Some of the different shapes of the beta distribution, depending on the values of the mean and dispersion parameters, can be visualized in Figure~\ref{grafdens}. 
  
Considering a vector of $n$ random variables $y_t$, $t=1,2,\ldots,n$, with mean $\mu_{t}$ and dispersion $\sigma_{t}$ parameters, and density given by \eqref{e:a}, the beta regression model with varying dispersion \citep{Cribari-NetoandSouza:2012, BayerandCribari:2014, Bayer2015} is defined by the following regression structures for $\mu_{t}$ and $\sigma_{t}$:  
\begin{align} g(\mu_{t})=&\sum_{i=1}^k x_{ti}\beta _{i}, \label{e:mean}\\
h(\sigma_{t})= &\sum_{i=1}^s z_{ti} \gamma_{i}, \label{e:disp}
\end{align}
where $\beta=(\beta_{1}, \ldots ,\beta_{k})^\top$ and $\gamma=(\gamma_{1}, \ldots ,\gamma_{s})^\top$ are vectors of unknown parameters, $x_{1t},\ldots,x_{kt}$ are the $k$ covariates of the mean submodel and $z_{1t},\ldots,z_{st}$ are the $s$ covariates of the dispersion submodel. When intercepts are included in the mean and dispersion submodels, we have $x_{1t}=z_{1t}=1$, for $t=1,\ldots,n$. Finally, $g(\cdot)$ and $h(\cdot)$ are the strictly monotonous and twice differentiable link functions, with domain in $(0,1)$ and image in $\R$. There are several possible choices for link functions such as logit, probit, log-log, complement log-log, Cauchy, and also parametric links \citep{Canterle2017}.

The purpose of BRCC is to monitor fraction or proportion type variables, for situations where the mean and the dispersion of the quality characteristic of interest are affected by control variables. 
Given an average run length under control (ARL$_0$) of interest, it is possible to determine 
$\alpha=\frac{1}{\operatorname{ARL}_0}$. 
Thus, the limits of the proposed control chart are defined by: 
\begin{align}
\mbox{UCL}_t&=Q(\alpha/2;\mu_{t},\sigma_{t}), \label{e:m}\\
\mbox{LCL}_t&=Q(1-\alpha/2;\mu_{t},\sigma_{t}),\label{e:n}
\end{align}  
where $\mu_{t}$ and $\sigma_{t}$ are given by the mean regression structures 
$\mu_t=g^{-1}(\sum_{i=1}^k x_{ti}\beta_i)$ in \eqref{e:mean} and of the dispersion 
$\sigma_t=h^{-1}(\sum_{i=1}^s z_{ti}\gamma_i)$ in \eqref{e:disp} and $\alpha$ is the fixed probability of false alarms.  
In practice, the MLE of $\mu_t$ and $\sigma_t$ are considered, with 
$\hat{\mu}_t=g^{-1}(\sum_{i=1}^k x_{ti} \hat{\beta}_i)$ and 
$\hat{\sigma}_t=h^{-1}(\sum_{i=1}^s z_{ti} \hat{\gamma}_i)$, 
where $\hat{\beta}$ and $\hat{\gamma}$ are the MLE of $\beta$ and $\gamma$, respectively.
The MLE of $\beta$ and $\gamma$ can be carried out by log-likelihood function maximization. The log-likelihood function is  
\begin{align}
\ell(\beta,\gamma)=\sum^n_{t=1}\ell_t(\mu_t,\sigma_t),
\end{align}
where  
\begin{align}
\ell_t(\mu_t,\sigma_t) &= \log \Gamma{\left( \frac{1-\sigma^2_t}{\sigma^2_t} \right)} \! - \! \log \Gamma{\left(\! \mu_t \left(\frac{1-\sigma^2_t}{\sigma^2_t}\right) \!\!\right)} \! - \! \log \Gamma{\left(\! (1-\mu_t) \!  \left(\frac{1-\sigma^2_t}{\sigma^2_t} \right) \!\! \right)}\\ \nonumber
& + \left[\mu_t \left(\frac{1-\sigma^2_t}{\sigma^2_t} \right)-1 \right] \log y_t + \left[(1-\mu_t)\left(\frac{1-\sigma^2_t}{\sigma^2_t}\right)-1 \right]\log(1-y_t). 
\end{align} 
Details on the score function, Fisher's information matrix, and large sample inferences in this model can be found in \cite{Cribari-NetoandSouza:2012}. For the explanatory variables selection, both in the mean and the dispersion structure, model selection criteria can be considered as widely discussed in \cite{BayerandCribari:2014, Bayer2015}. 
Residuals and other diagnostic tools in beta regression models are discussed in \cite{FerrariandCribari:2004} and \cite{Espinheiraetal:2008,espinheira2008}.

It is noteworthy that it is possible to test whether dispersion is constant, i.e., test the null hypothesis $$\mathcal{H}_0 : \sigma_1 = \sigma_2 = \cdots = \sigma_n = \sigma. $$ Equivalently, we can test $$ \mathcal{H}_0 : \gamma_i = 0,\quad  i=2,\ldots,s, $$ for the model given in \eqref{e:disp} with $z_{t1}=1$, $t=1,\ldots,n$. The likelihood ratio statistic is given by: 
\begin{align} \label{e:lr}
{\rm LR}=2 
\left[
\ell (\hat{\beta},\hat{\gamma})
-
\ell (\tilde{\beta},\tilde{\gamma})
\right],
\end{align}
where $\hat{\beta}$ and $\hat{\gamma}$ are the unrestricted MLE (under alternative hypothesis) and $\tilde{\beta}$ and $\tilde{\gamma}$ are the restricted MLE (under $\mathcal{H}_0$). Under the usual regularity conditions and under $\mathcal{H}_0$, ${\rm LR}$ converges in distribution to $\chi^2_{(s-1)}$. The null hypothesis of constant dispersion is thus rejected if ${\rm LR} > \chi^2_{1-a, s-1}$, where $\chi^2_{1-a, s-1}$ is the $1-a$ $\chi^2_{s-1}$ upper quantile, where $a$ is the test nominal level.

In this way, we propose the following algorithm to implement the beta regression control chart: 
\begin{enumerate}
\item Fit the beta regression model, obtaining the MLE of the parametric vectors $\beta$ and $\gamma$, $\hat{\beta}$ and $\hat{\gamma}$;
\item Compute the estimative of the mean, $\mu$, and dispersion, $\sigma$, for each $t$, with $t=1, \ldots,n$, given by 
$\hat{\mu}_t=g^{-1}(\sum_{i=1}^k x_{ti} \hat{\beta}_i)$ and 
$\hat{\sigma}_t=h^{-1}(\sum_{i=1}^s z_{ti} \hat{\gamma}_i)$, respectively. 
\item Determine the estimated control limits, for a given ARL$_0$, by:
\begin{align}
\widehat{\mbox{UCL}}_t&=Q(\alpha/2;\hat{\mu}_{t},\hat{\sigma}_{t}), \\
\widehat{\mbox{LCL}}_t&=Q(1-\alpha/2;\hat{\mu}_{t},\hat{\sigma}_{t}),
\end{align}  
where $\alpha=\frac{1}{\operatorname{ARL}_0}$.
\item Each data point $y_t$ is plotted together with the estimated control limits $\widehat{\mbox{UCL}}_t$ and $\widehat{\mbox{LCL}}_t$, with $t=1, \ldots,n$.
\end{enumerate}
The observation $y_t$ that is out of the control limits interval ($\widehat{\mbox{UCL}}_t, \widehat{\mbox{LCL}}_t)$ is considered out-of-control.

The main advantage of the beta regression control chart over traditional regression control charts is that it naturally accommodates asymmetric and heteroscedastic variables, and its limits are restricted to the interval $(0,1)$. Moreover, it allows for the modeling of one structure for $\mu_{t}$ and one for $\sigma_{t}$ simultaneously, and the assumption of non-constant mean and dispersion is natural in several production processes \citep{Gan1995, Riaz2008, Sheu2009}. In practical situations, it is important to monitor changes in the dispersion of the process, because an increase in dispersion may indicate process deterioration, while a reduction in dispersion means an improvement in the process capability \citep{Huwangetal:2010}. 

\section{Numerical evaluation}\label{S:resultados} 

This section presents results of Monte Carlo simulations for numerical evaluation of the beta regression control chart in its versions with constant dispersion (BRCC$_C$) and variable dispersion (BRCC), compared to the traditional regression control chart (RCC) \citep{Mandel:1969}. The performance evaluation of the control charts will be performed computing the ARL \citep{Montgomery:2009}. For this, a process under control and another out-of-control are evaluated. For the process under control, we evaluate ARL$_{0}$, that is, the average of observations until a false special cause is detected. A higher ARL$_{0}$ indicates a lower probability of false alarms \citep{CastagliolaandMaravelakis:2011, Montgomery:2009}. On the other hand, for a fixed ARL$_{0}$, considering an out-of-control process, we have ARL$_{1}$, which evaluates the average run length of observations until a true special cause is detected. A smaller ARL$_1$ indicates a lower average number of samples collected until the introduced change in the process is detected \citep{SantAnnaaandCaten:2012}. 
ARL$_{0}$ and ARL$_{1}$ are defined, respectively, as follows \citep{Montgomery:2009}: 
\begin{align}
{\rm ARL}_0&=\frac{1}{\hat{\alpha}}, \\  
{\rm ARL}_1 &=\frac{1}{[1-\hat{\beta}]},
\end{align}
where $\hat{\alpha}$ is the probability of false alarm (type I error) and $\hat{\beta}$ is the probability of false control (type II error). Mathematically, assuming $y\sim Beta(\mu,\sigma)$, the errors can be denoted by:  
\begin{align*}
\hat{\alpha}&= P(y \not\in [LIC,LSC]|\mu=\mu_0 \quad \text{and} \quad \sigma = \sigma_0),  \\
\hat{\beta}&= P(y \in [LIC,LSC]|\mu=\mu_1 \quad \text{or} \quad  \sigma= \sigma_1), \quad \text{ with } \quad \mu_1 = \mu_0 +\delta_\mu \quad \text{and} \quad \sigma_1 = \sigma_0 +\delta_\sigma,
\end{align*}
where $\mu_0$ is the average of the process under control, $\sigma_0$ is the dispersion of the process under control, $\mu_1$ is the average of the out-of-control process, $\sigma_1$ is the dispersion of the out-of-control process, and $\delta_\mu$ and $\delta_\sigma$ are the changes induced in the process of computing ARL$_1$. 


For the numerical evaluation were considered $50000$ Monte Carlo replications. The sample sizes evaluated were $n = 200$, $300$, $500$ e $1000$. However, due to similarity of results and for brevity, only the results for $n = 200$ and $n = 1000$ will be presented here. The computational implementations were developed in $\tt{R}$ \citep{R2016} language. For the estimation of the beta regression model parameters, the $\tt{GAMLSS}$ \citep{Rigby2005} package was used. The generation of the data under control was based on the beta regression model with structures for the mean and the dispersion given, respectively, by: 
\begin{align}
g(\mu_{t})&=\beta_{0}+\beta_{1}x_{t1}+\beta_{2}x_{t2},\\
h(\sigma_{t})&= \gamma_{0}+\gamma_{1}z_{t1}+\gamma_{2}z_{t2}, 
\end{align}
where $g(\mu)=\operatorname{logit}(\mu)=\log\left(\frac{\mu}{1-\mu}\right)$, $h(\sigma)=\operatorname{logit}(\sigma)=\log\left(\frac{\sigma}{1-\sigma}\right)$ and the values of the covariates are values from the standard uniform distribution $\mathcal{U}(0,1)$, kept constant during the whole experiment. For each Monte Carlo replication, a $y_1,\ldots,y_n$ sampling with beta density is generated, with beta density given by \eqref{e:a}. The parameter values for the mean and dispersion structures are presented in Table \ref{cenario}. The scenarios were created to address different possible characteristics of the data on the $(0,1)$ range. For Scenarios $1$, $2$, and $3$,  the approximate value of the mean is $\mu \cong 0.4$ and the values of $\gamma$ vary in order to take into account different levels of dispersion. Scenarios $4$, $5$, and $6$ have mean values around to $\mu \cong 0.2$, $\mu\cong0.8$ and $\mu \cong 0.08$, respectively, with a small dispersion of $\sigma \cong 0.070$. 

\begin{table}[t] 
\small
\caption{Parameter values for each scenario considered in the Monte Carlo simulation.}
\label{cenario}
\begin{center}
\begin{tabular}{c|cccccc|c}
\hline 
Scenario& $\beta_{0}$ & $\beta_{1}$& $\beta_{2}$  & $\gamma_{0}$ & $\gamma_{1}$ & $\gamma_{2}$ & Characteristic \\
\hline
1& -1.35& 1.00 & 1.00&   -1.40 & 1.00&-1.25 &   $\mu \cong 0.40 \quad \text{and} \quad \sigma \cong 0.156$ \\
2& -1.35& 1.00 & 1.00&  -1.00 & -1.10&-1.00 &   $\mu \cong 0.40 \quad \text{and} \quad \sigma \cong 0.099$ \\
3& -1.35& 1.00 & 1.00&  -1.15 & -1.20 & -1.20 &   $\mu \cong 0.40 \quad \text{and} \quad \sigma \cong0.070$ \\
4& -0.10 & -1.35& -1.40 & -1.15 & -1.20 & -1.20 & $\mu \cong 0.20 \quad \text{and} \quad \sigma \cong0.070$ \\ 
5& 1.50 & 1.00& -1.00 & -1.15 & -1.20 & -1.20 &   $\mu \cong 0.80 \quad \text{and} \quad \sigma \cong 0.070$ \\
6& -1.00 & -1.50& -1.50 & -1.15 & -1.20 & -1.20 &   $\mu \cong 0.08 \quad \text{and} \quad \sigma \cong 0.070$ \\
\hline
\end{tabular} 
\end{center} 
\end{table}

The performance of the control charts was evaluated by accessing the ARL values. In order to evaluate ARL$_1$, we first carried out a simulation study to compare the control charts in terms of ARL$_0$, as performed in \cite{CapizziandMasarotto:2011}, \cite{Moraes2014} and \cite{Zhangetal2015}. In all experiments, ARL$_0 = 200$ was kept constant. 
To compute ARL$_1$ we considered two situations, namely: (i) changes in the mean process and (ii) changes in the dispersion of the process.  
The change $\delta$  in the mean process was induced as follows: $g(\mu_t)=\delta+\beta_0+\beta_1x_{t1}+\beta_2x_{t2}$. The values $\delta$ that were used varied from $-0.15$ to $0.15$, in increments (steps) of $0.01$, representing different magnitudes of change. 
In the same way, 
the change $\delta$  in the dispersion process was considered as follows: $h(\sigma_t)=\delta+\gamma_0+\gamma_1z_{t1}+\gamma_2z_{t2}$, with $\delta$ ranging from $0.00$ to $0.15$, by steps of $0.01$. 
In the practical sense, an increase in dispersion may indicate process deterioration (out-of-control), while a reduction in dispersion means an improvement in the process capability \citep{Huwangetal:2010}.

\begin{figure} 
\begin{center} 
\subfigure[Scenario 1]{\includegraphics [width=8.0cm]{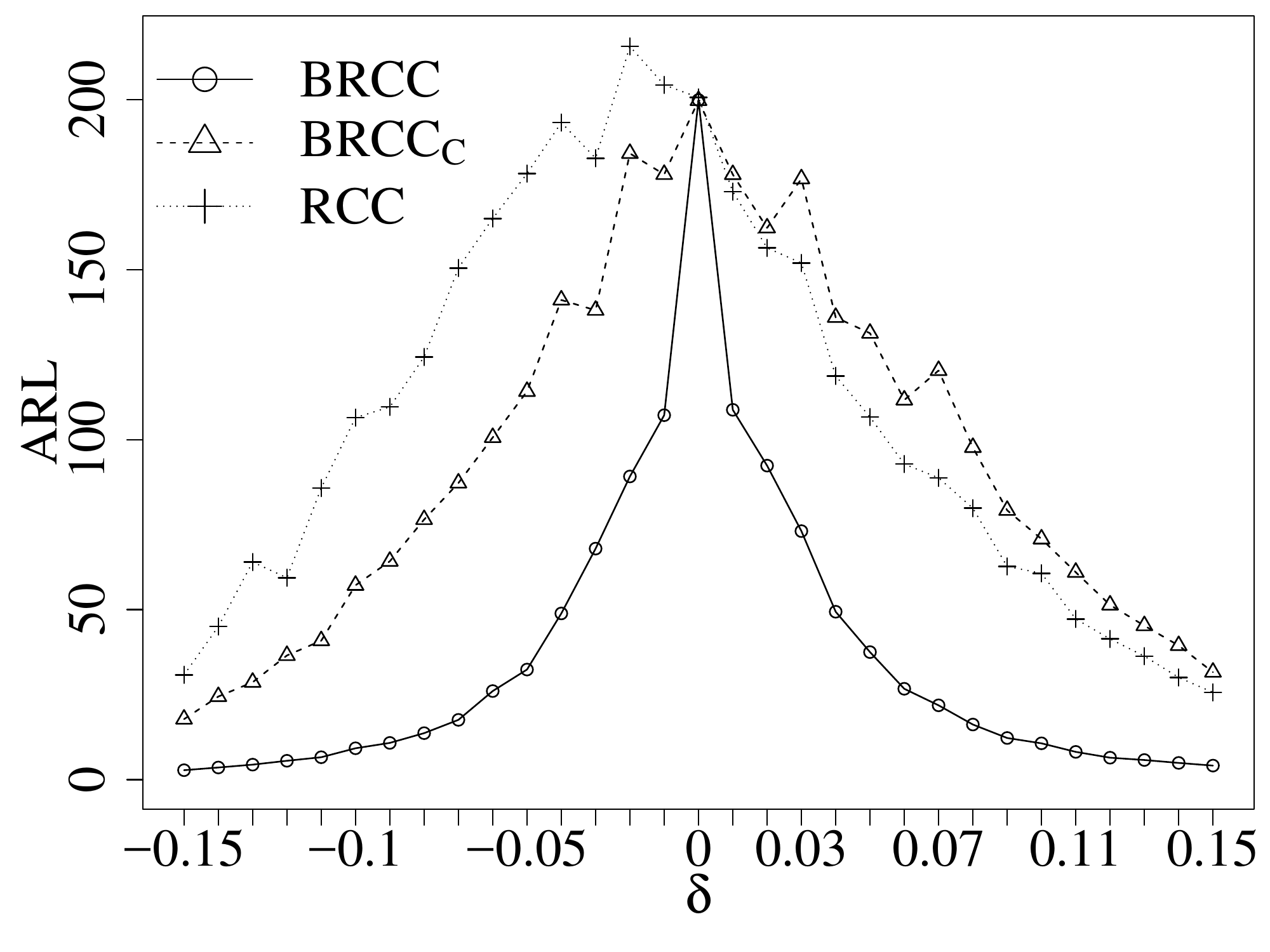}\label{f:1}} 
\subfigure[Scenario 2]{\includegraphics [width=8.0cm]{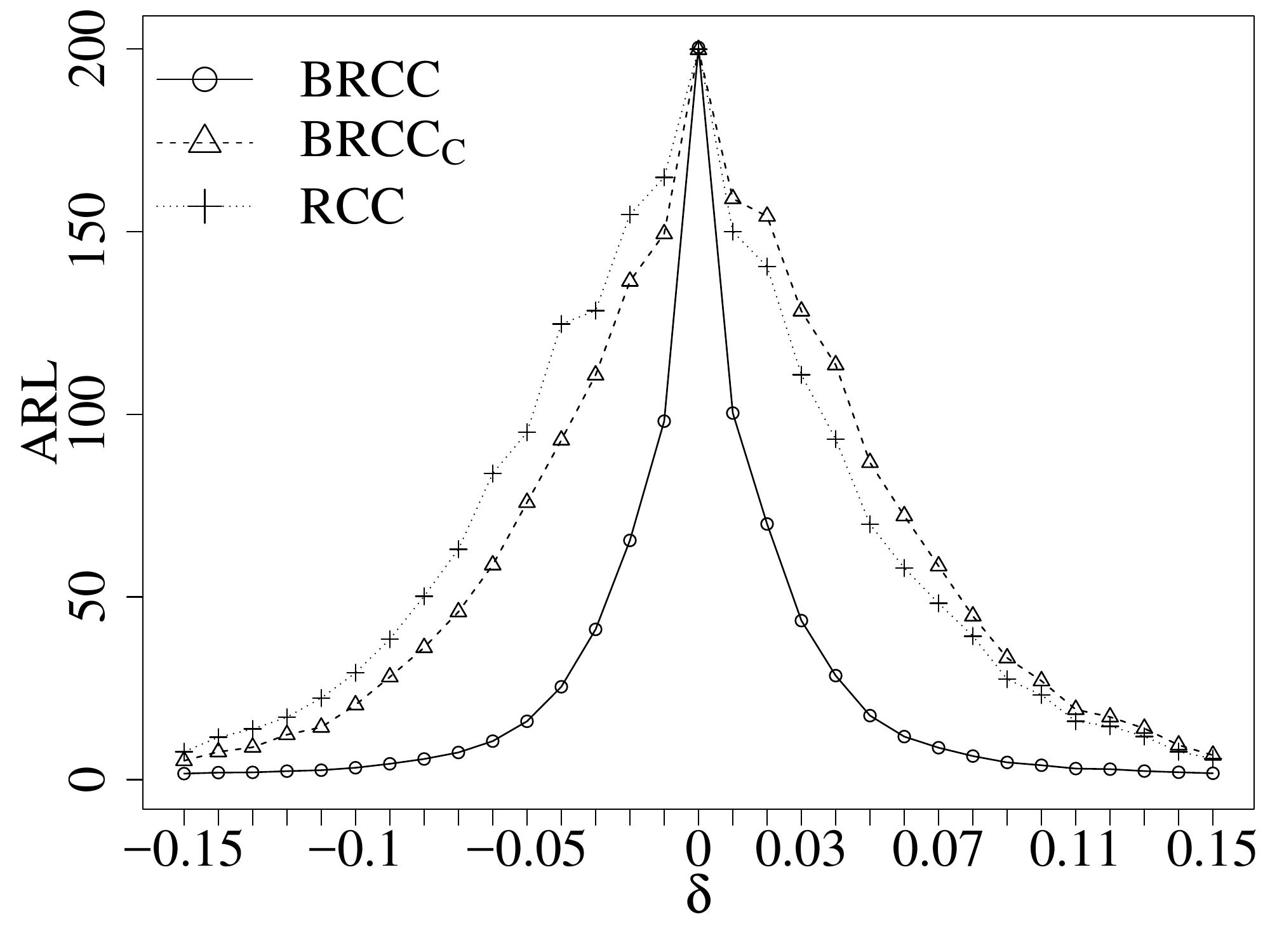}\label{f:2}} 
\subfigure[Scenario 3]{\includegraphics [width=8.0cm]{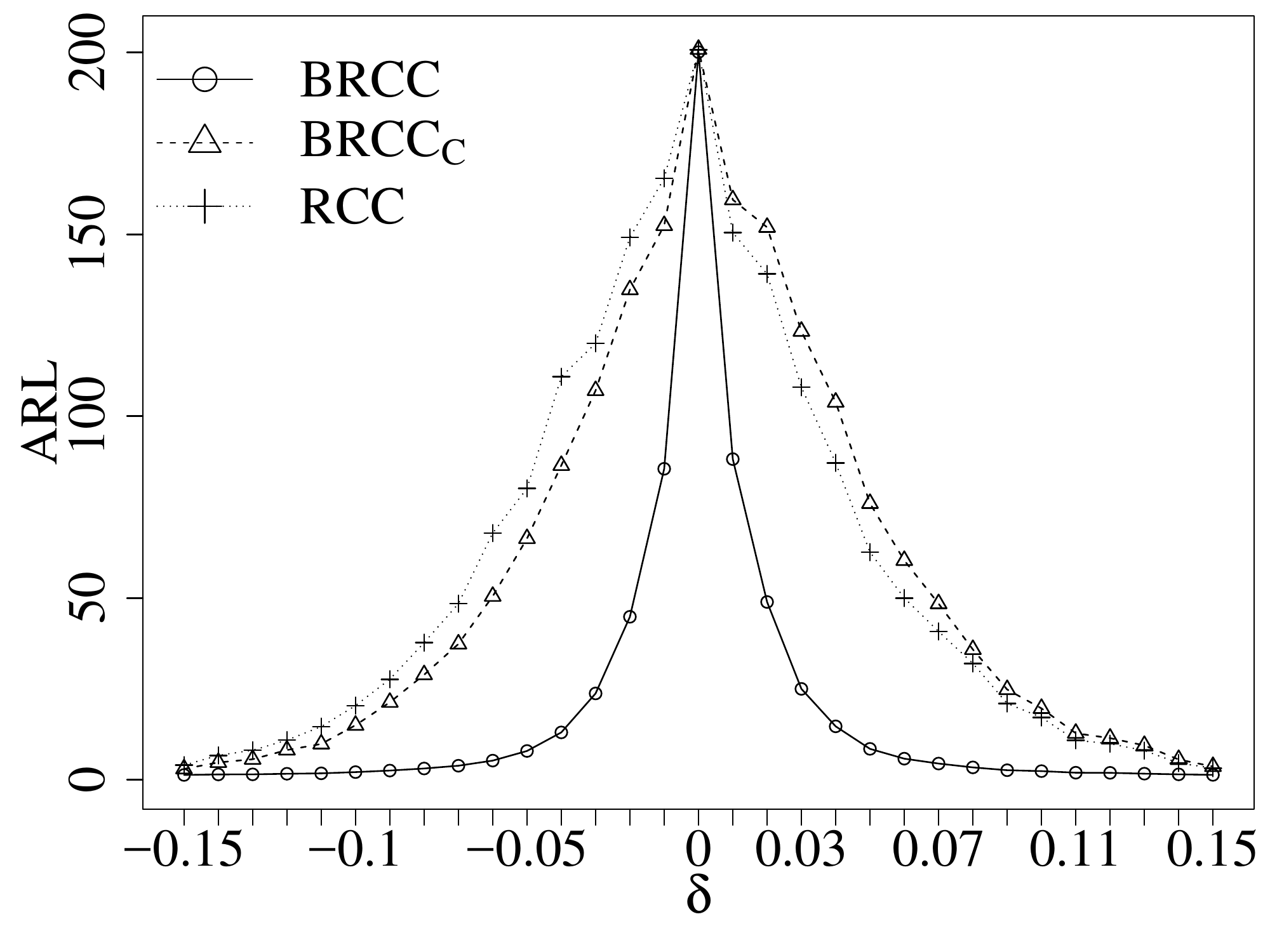}\label{f:3}} 
\subfigure[Scenario 4]{\includegraphics [width=8.0cm]{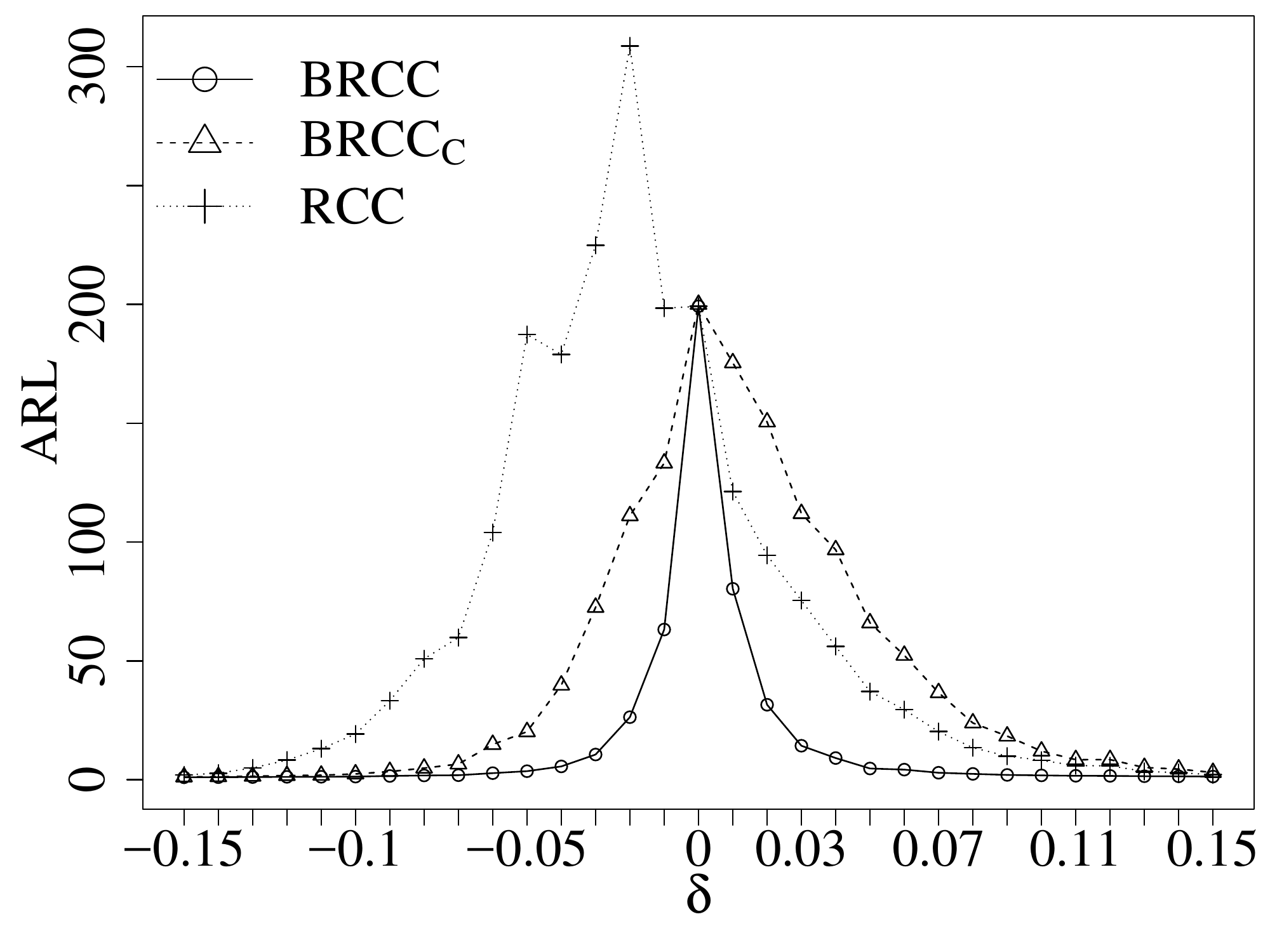}\label{f:4}} 
\subfigure[Scenario 5]{\includegraphics [width=8.0cm]{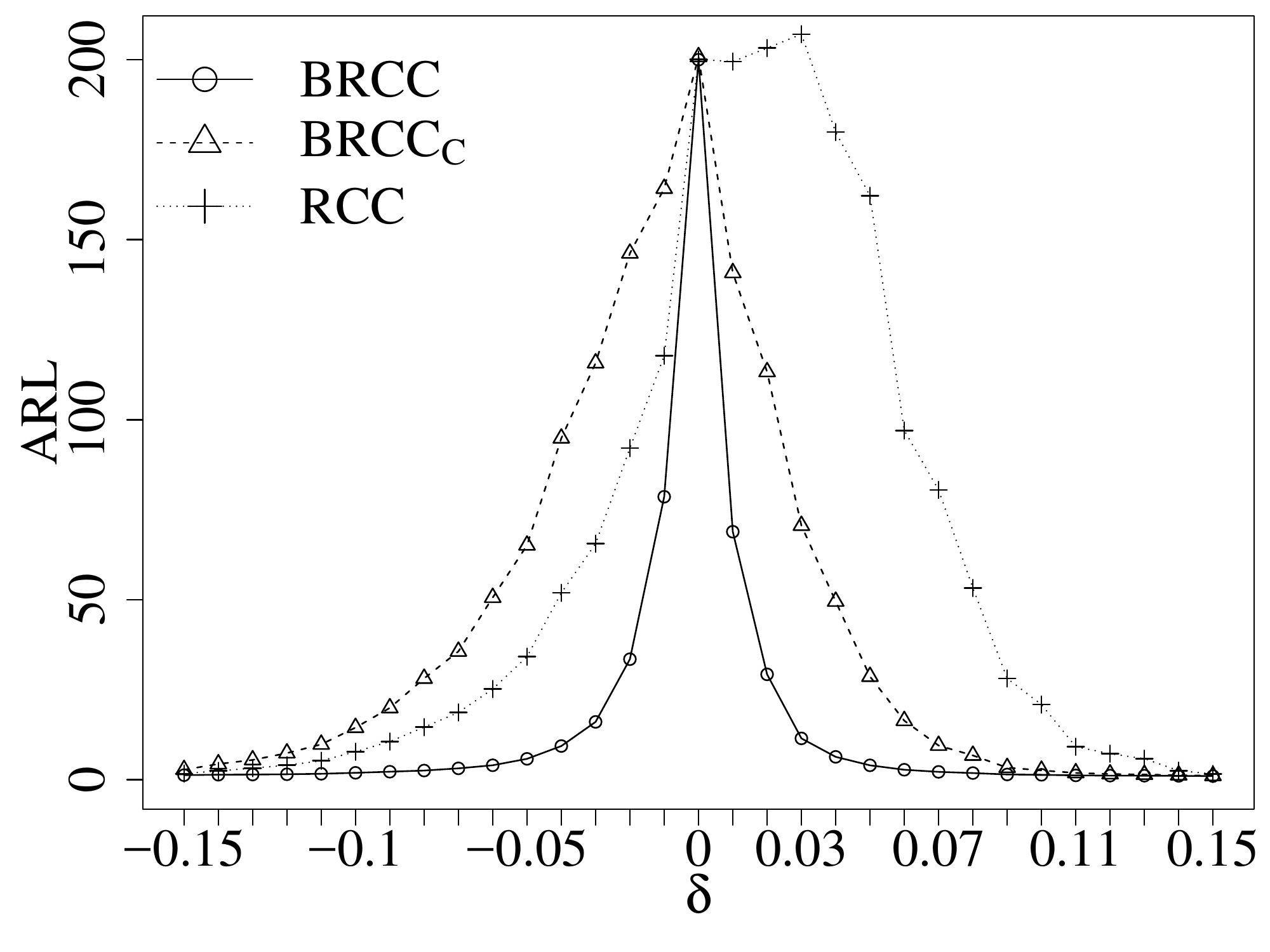}\label{f:5}}
\subfigure[Scenario 6]{\includegraphics [width=8.0cm]{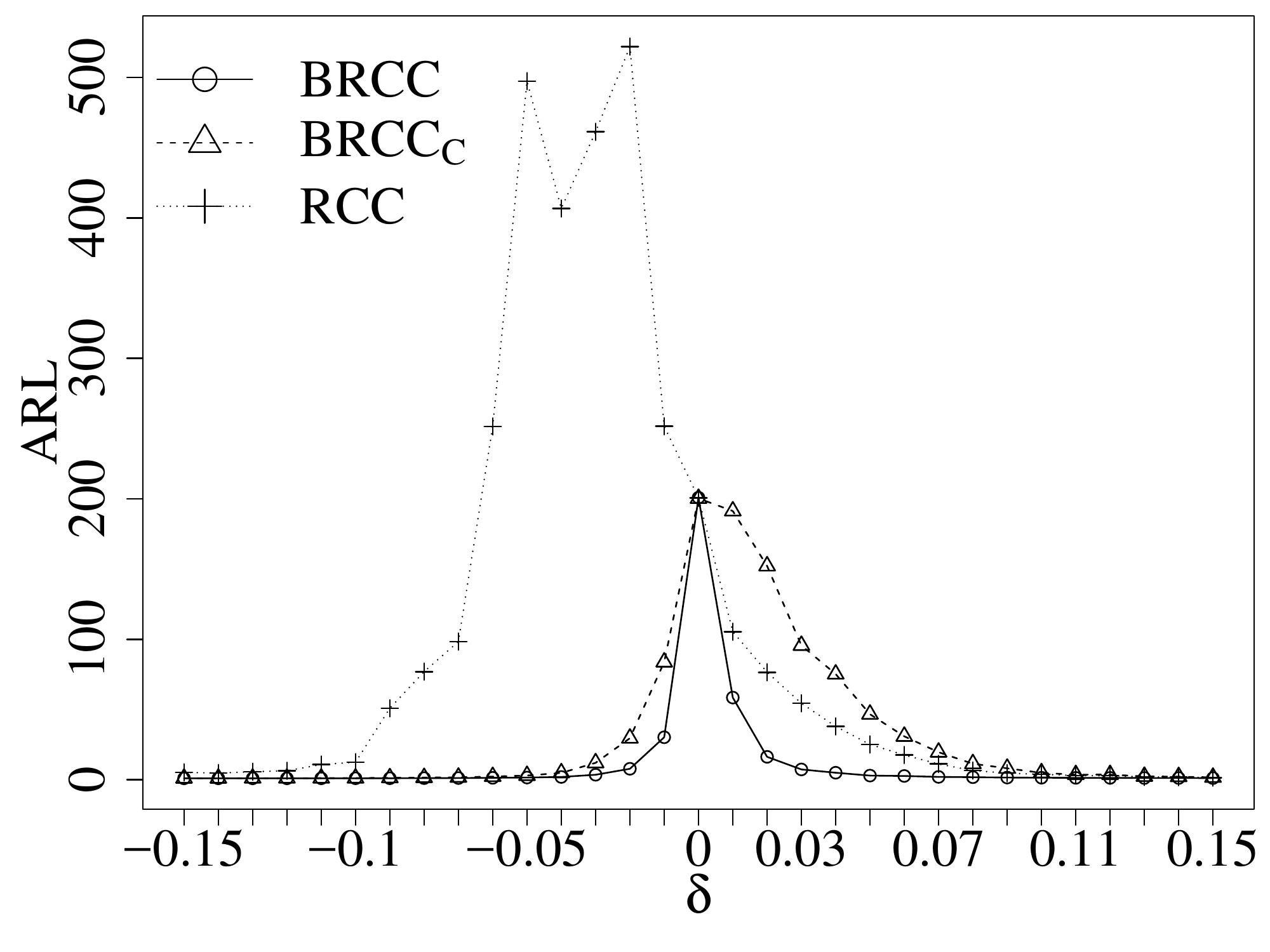}\label{f:6}}  
\caption{ARL$_1$ when the mean process is out-of-control, considering $n=200$.} 
\label{f:ARL200} 
\end{center} 
\end{figure}

\begin{figure}
\begin{center} 
\subfigure[Scenario 1]{\includegraphics [width=8.0cm]{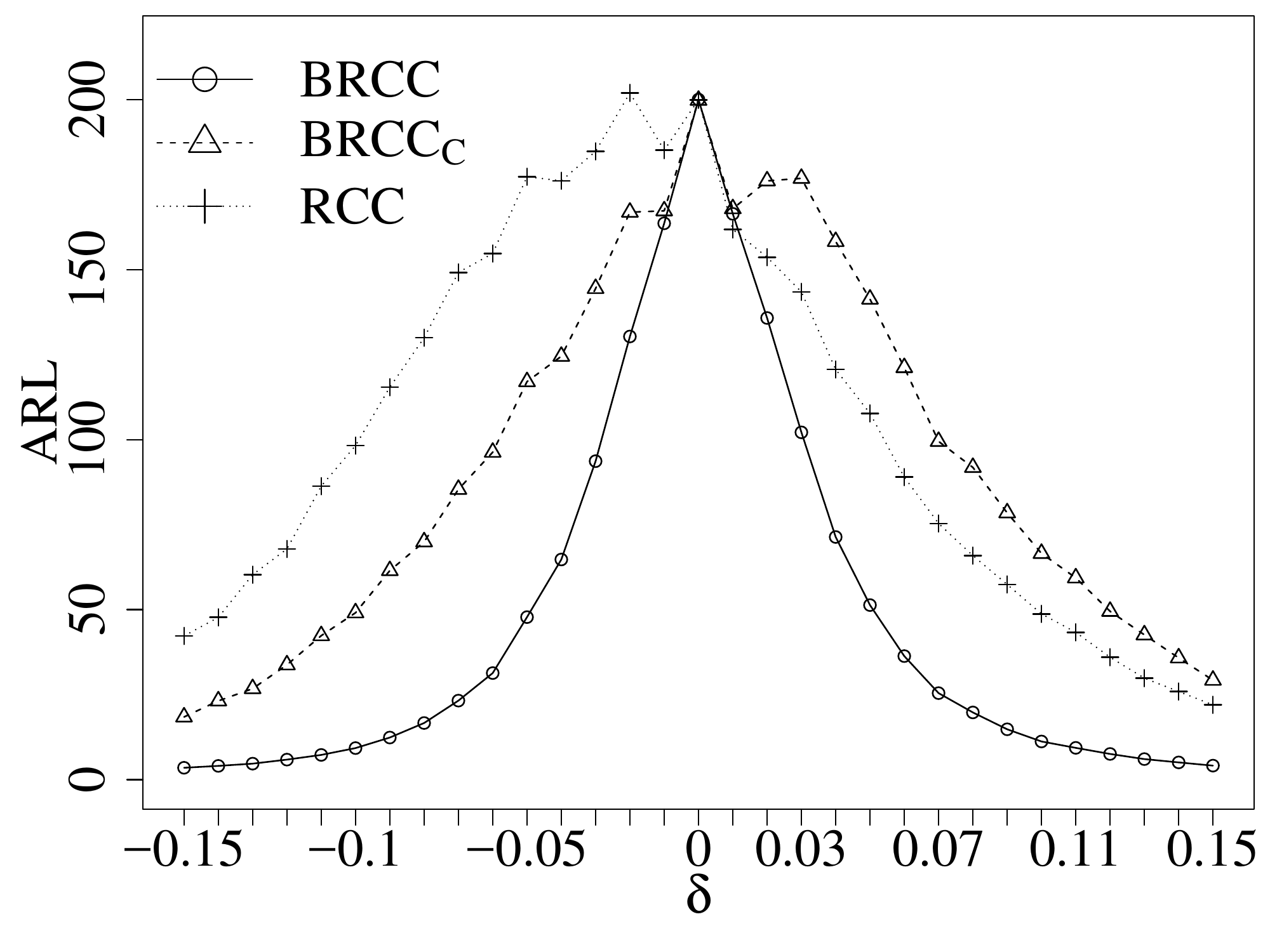}\label{f:1b}} 
\subfigure[Scenario 2]{\includegraphics [width=8.0cm]{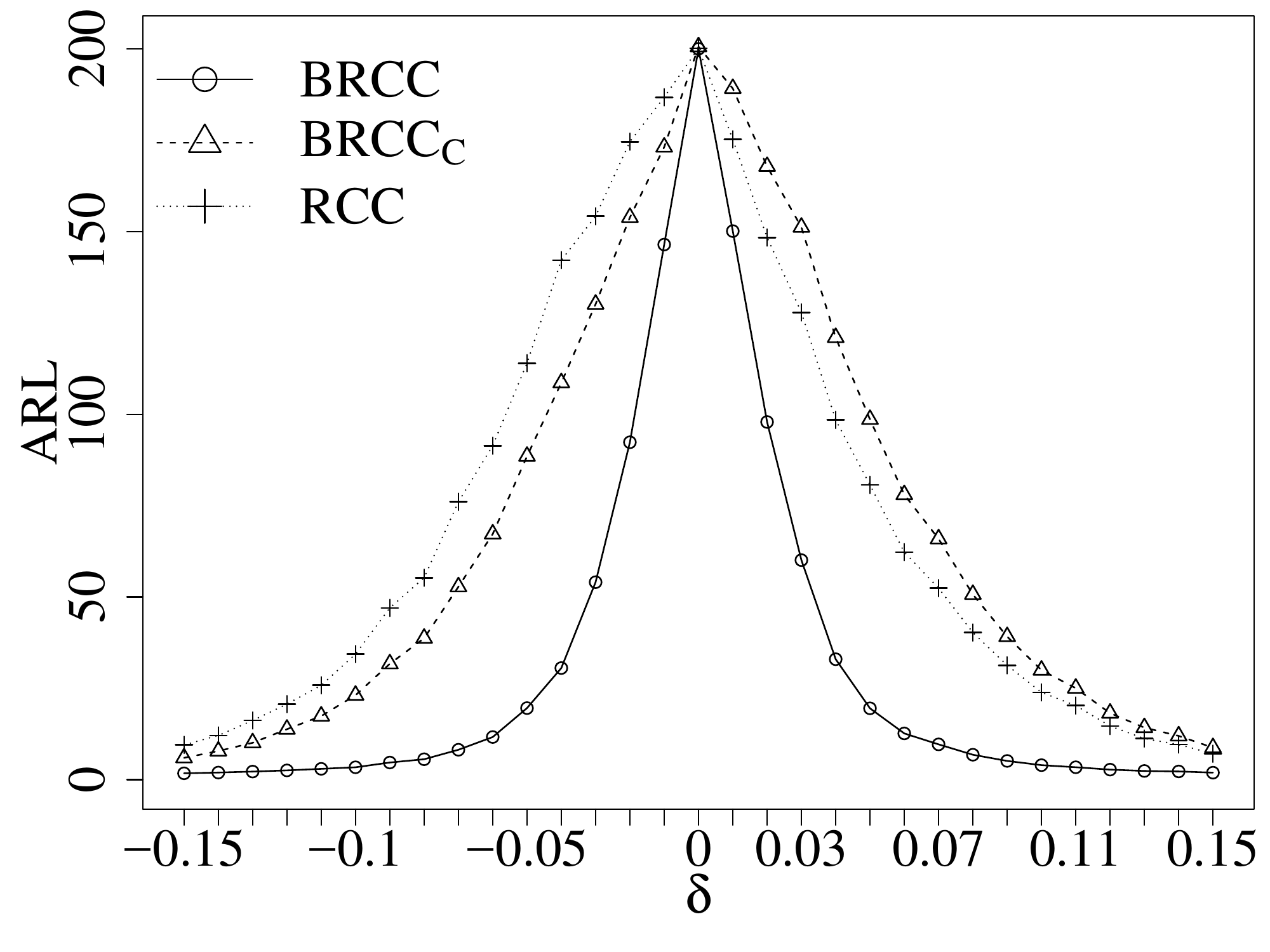}\label{f:2b}} 
\subfigure[Scenario 3]{\includegraphics [width=8.0cm]{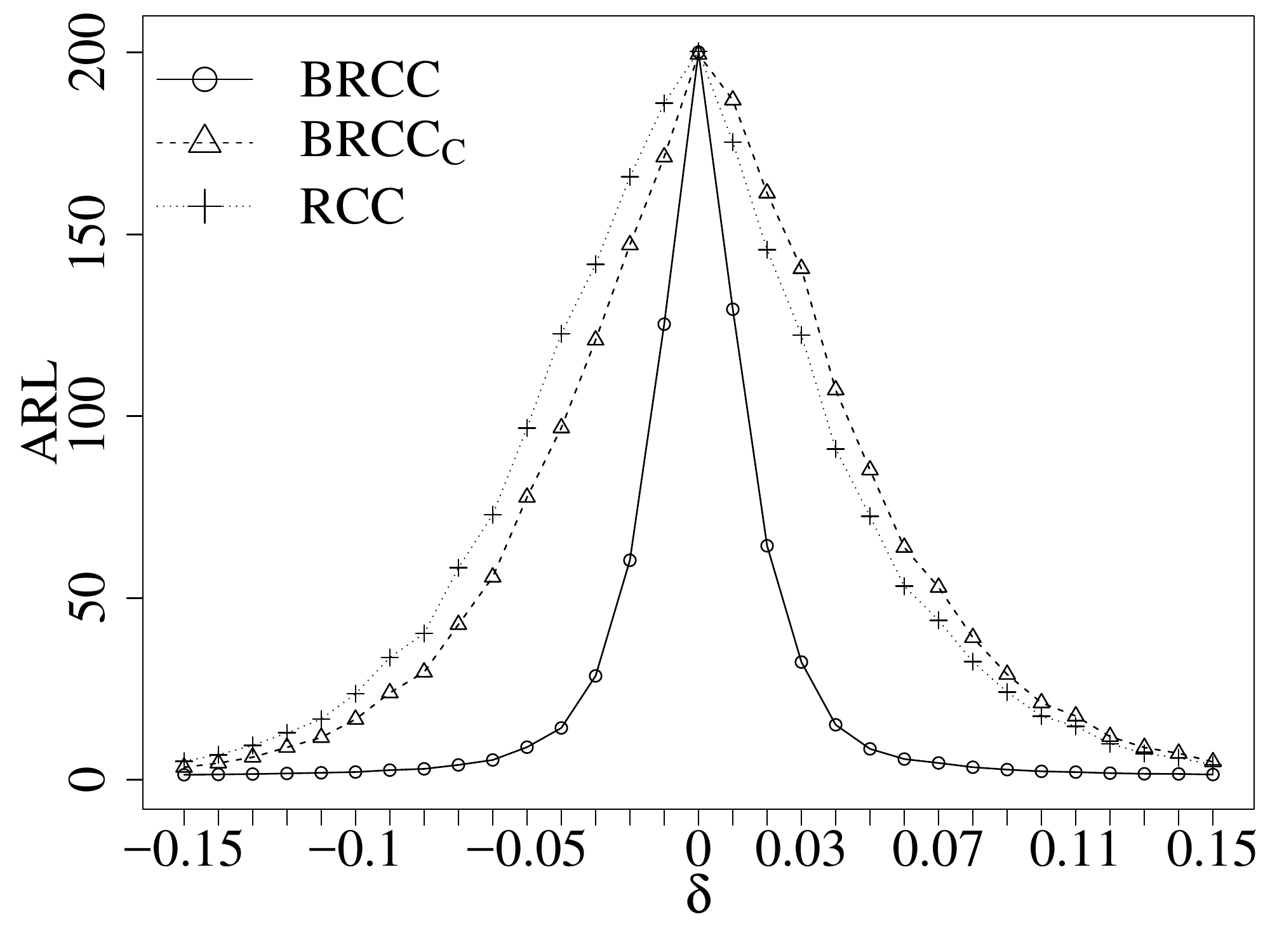}\label{f:3b}} 
\subfigure[Scenario 4]{\includegraphics [width=8.0cm]{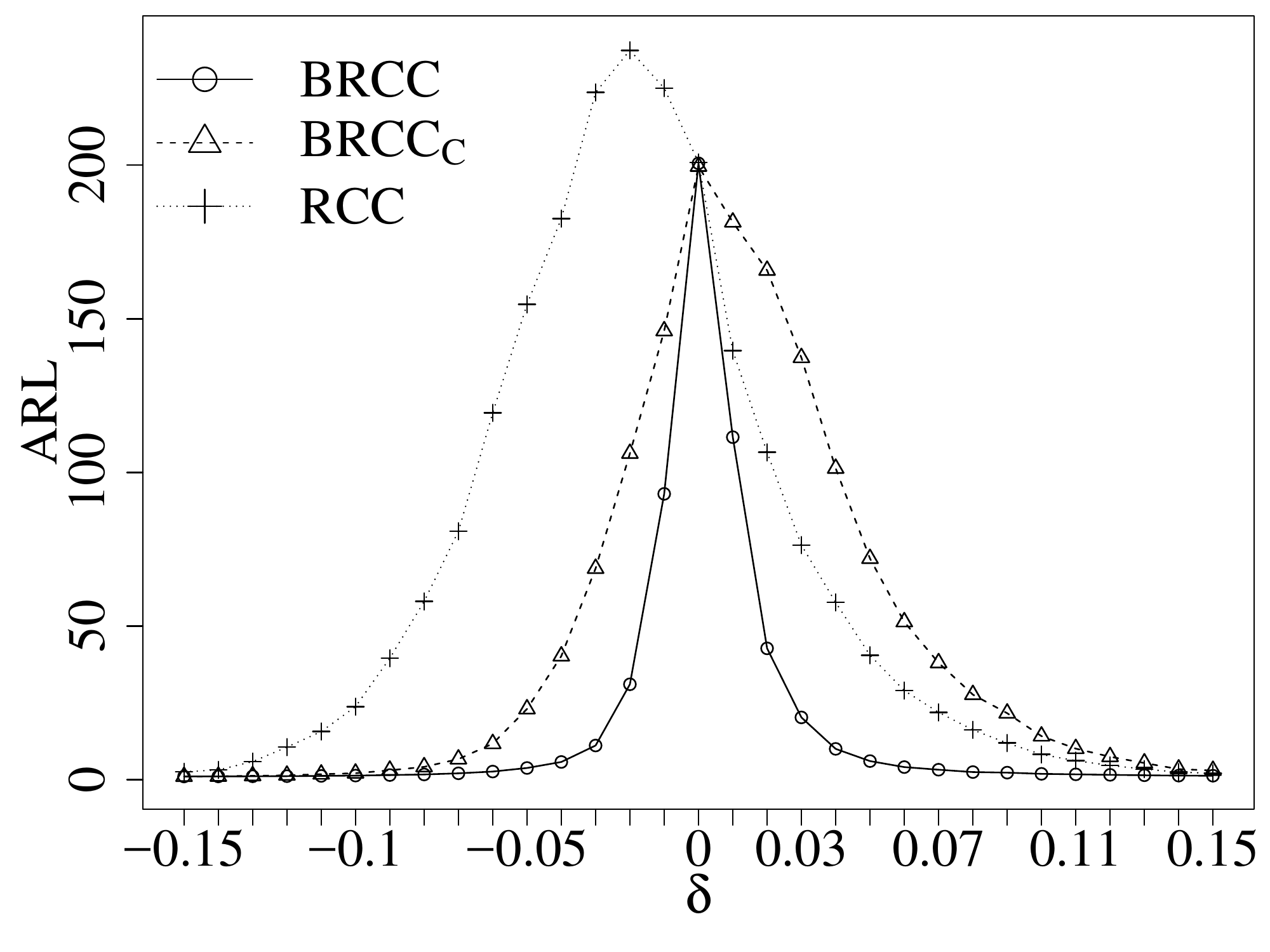}\label{f:4b}} 
\subfigure[Scenario 5]{\includegraphics [width=8.0cm]{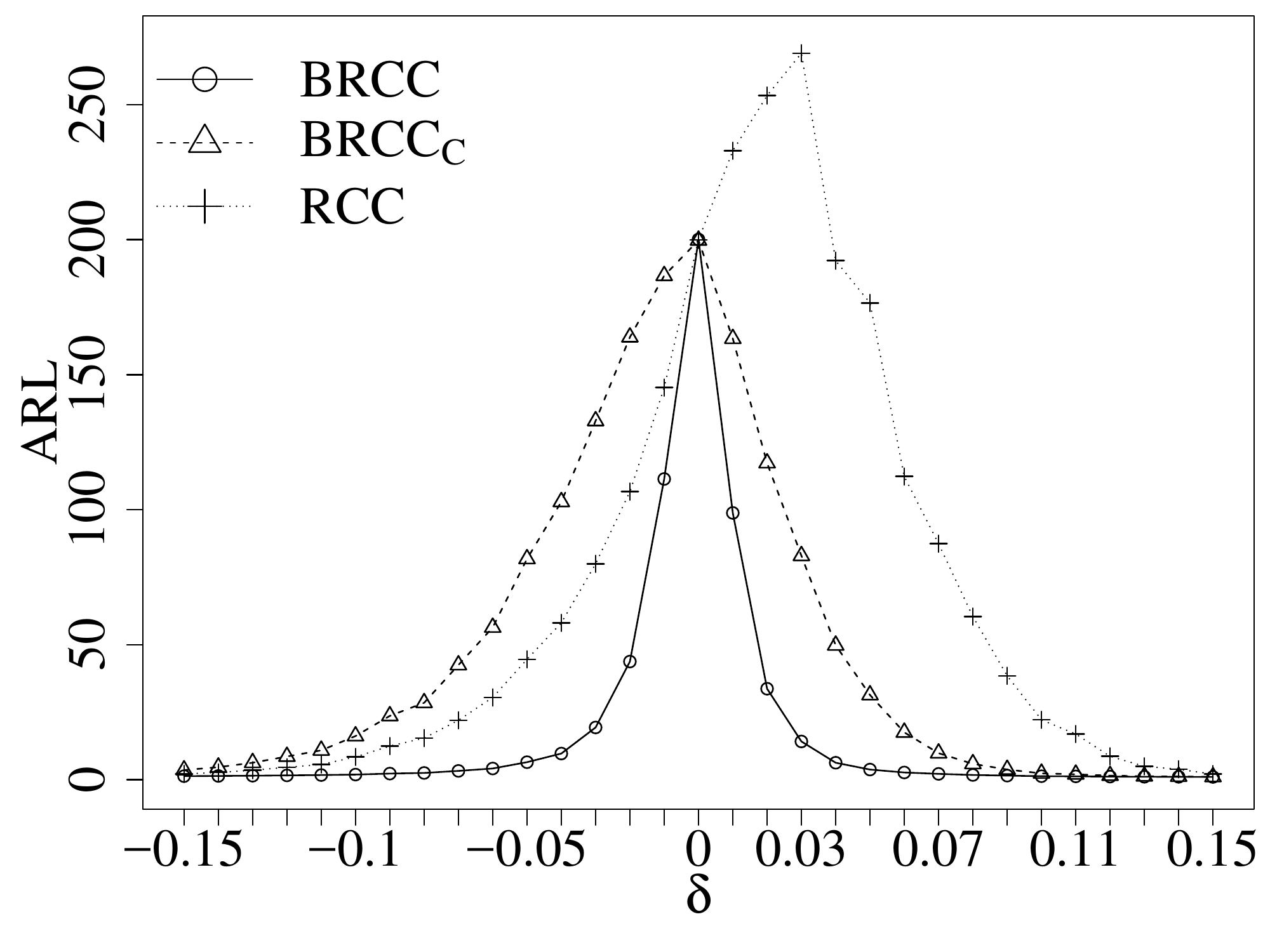}\label{f:5b}}
\subfigure[Scenario 6]{\includegraphics [width=8.0cm]{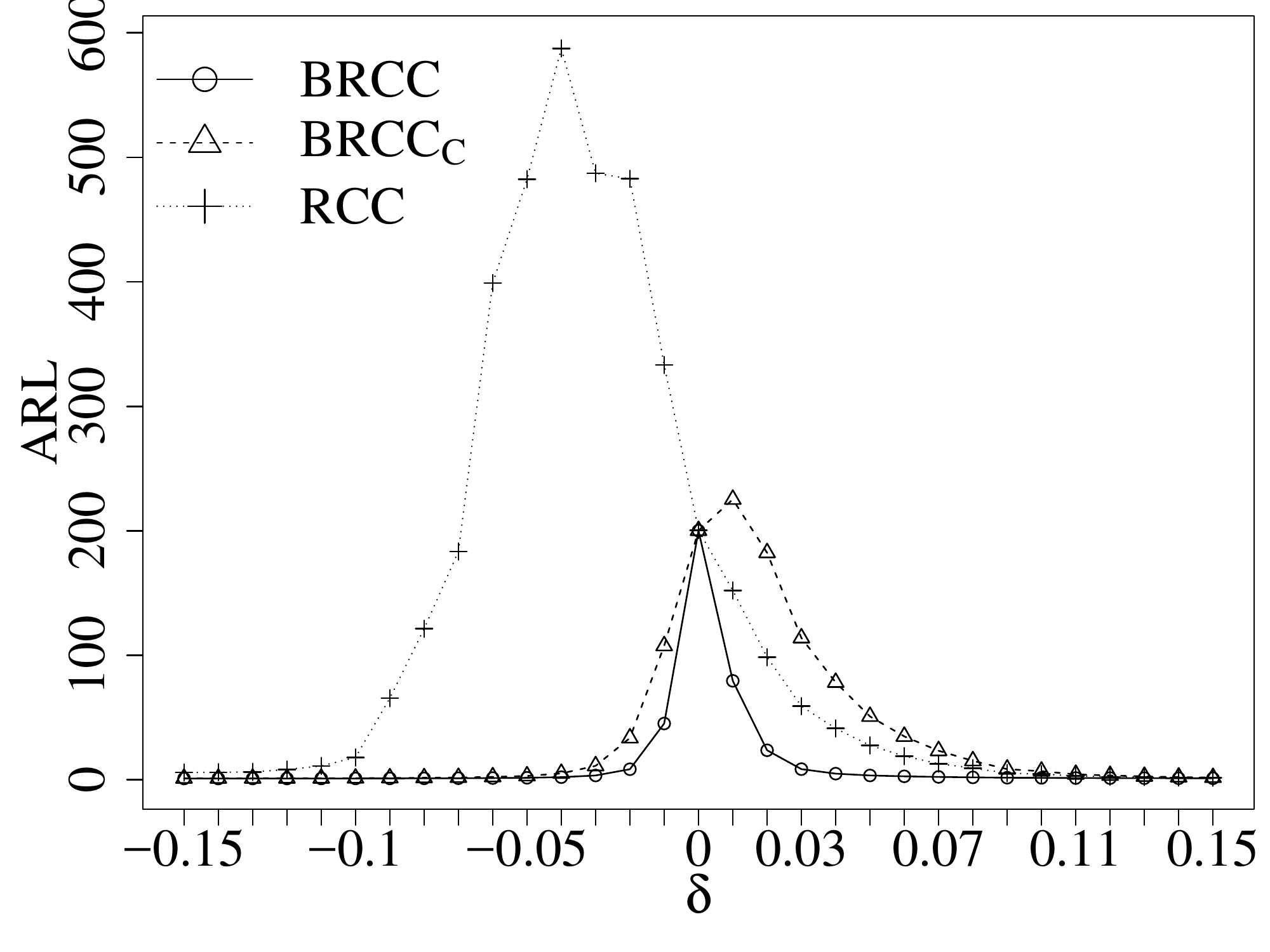}\label{f:6b}}  
\caption{ARL$_1$ when the mean process is out-of-control, considering $n=1000$.} \label{f:ARL1000} 
\end{center} 
\end{figure}

Figures \ref{f:ARL200} and \ref{f:ARL1000} present the results of ARL$_1$ for $n=200$ and $n=1000$, respectively, 
when the mean process is out-of-control.
When we analyze the numeric results of ARL$_1$ for the different scenarios (Figures \ref{f:ARL200} and \ref{f:ARL1000}), we observed that the control charts based on the beta regression model have a better performance when compared to RCC. For all levels of mean changes introduced, BRCC detects faster that the process is out-of-control. In Figure \ref{f:1}, Scenario $1$ for $n$ = 200 and $ \sigma\cong 0.156$, when the process $\delta=0$ is under control for the three charts, so we have ARL$_0 = 200$. As changes in the data generating process are inserted, the proposed control chart has a rapid decay, that is, it detects the changes more effectively than the other alternatives. For the first level of change, in Figure \ref{f:1}, the proposed control chart has $ARL\approx  100$, while for the BRCC$_C$ and RCC the decay is slower. This delays the detection of out-of-control processes. 

\begin{figure} 
\begin{center} 
\subfigure[Scenario 1]{\includegraphics [width=8.0cm]{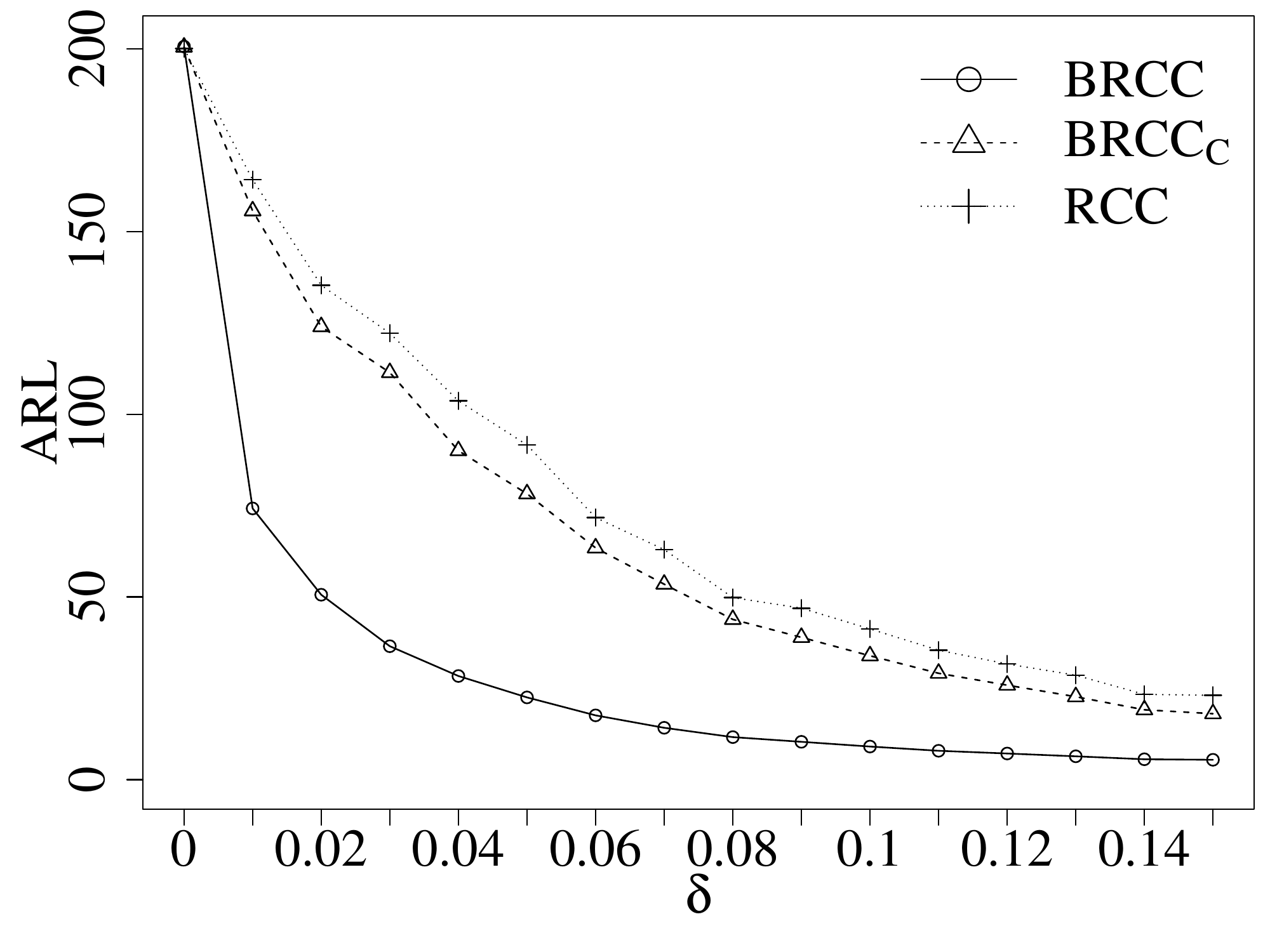}\label{f:1s}} 
\subfigure[Scenario 2]{\includegraphics [width=8.0cm]{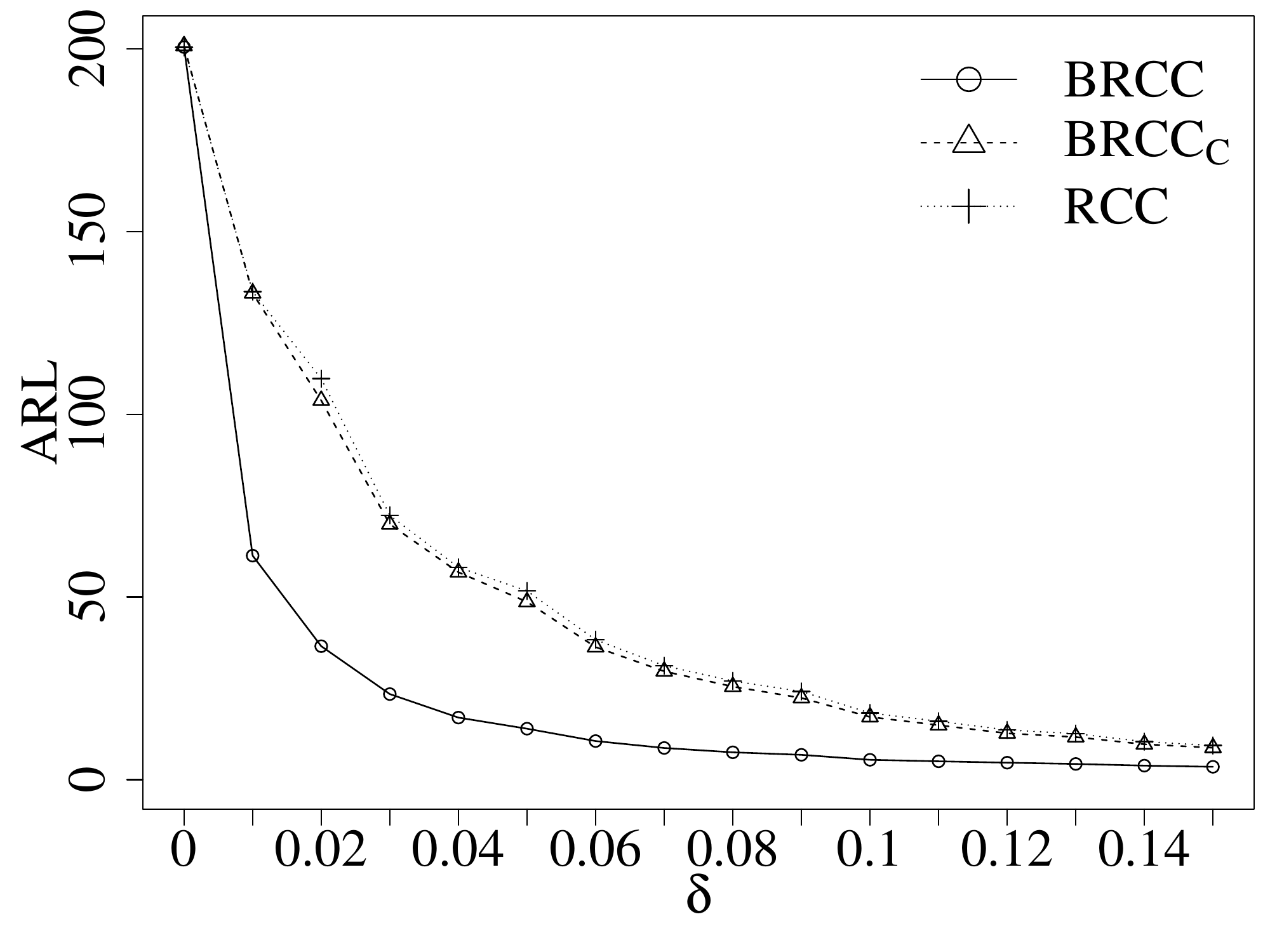}\label{f:2s}} 
\subfigure[Scenario 3]{\includegraphics [width=8.0cm]{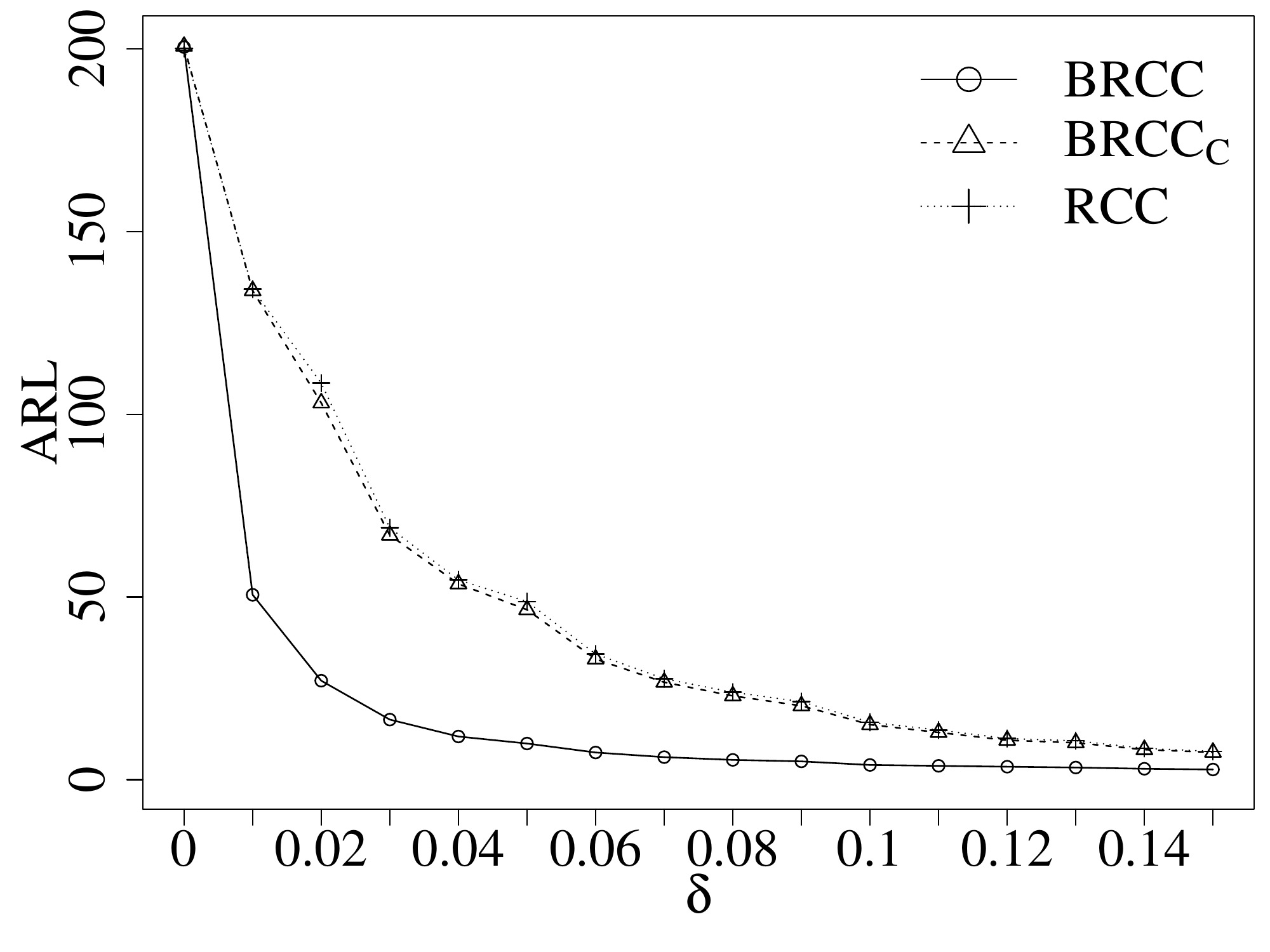}\label{f:3s}} 
\subfigure[Scenario 4]{\includegraphics [width=8.0cm]{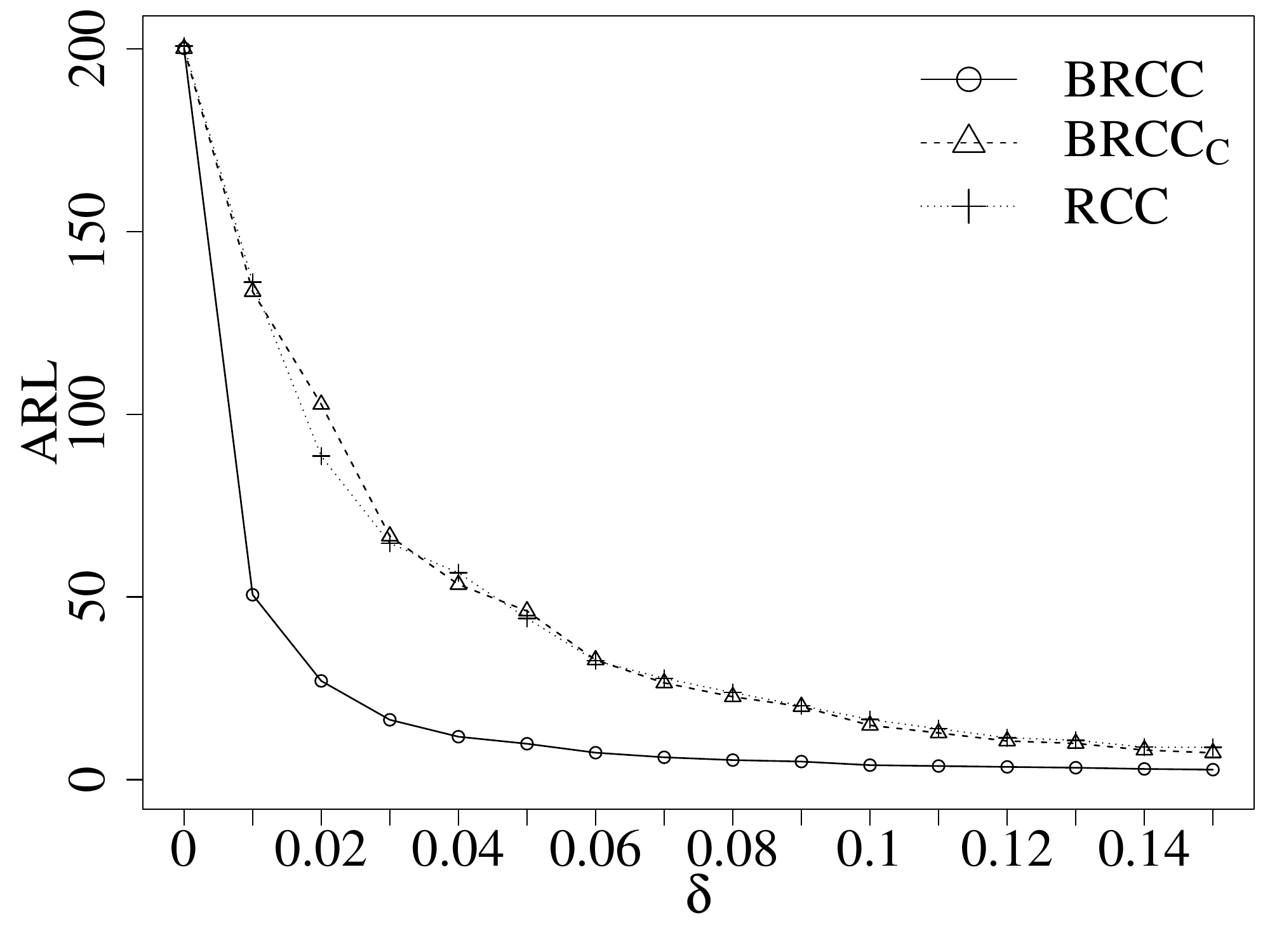}\label{f:4s}} 
\subfigure[Scenario 5]{\includegraphics [width=8.0cm]{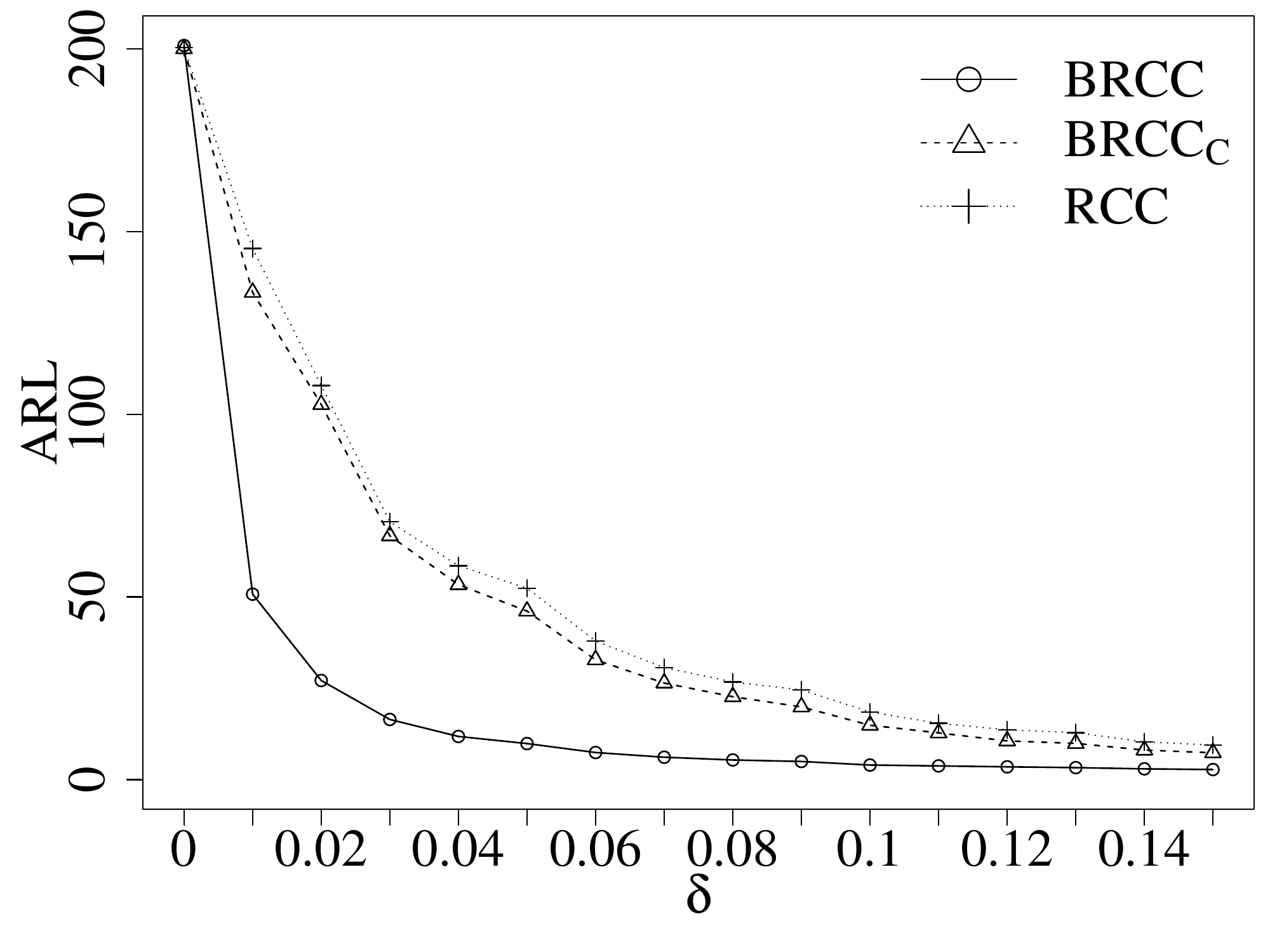}\label{f:5s}}
\subfigure[Scenario 6]{\includegraphics [width=8.0cm]{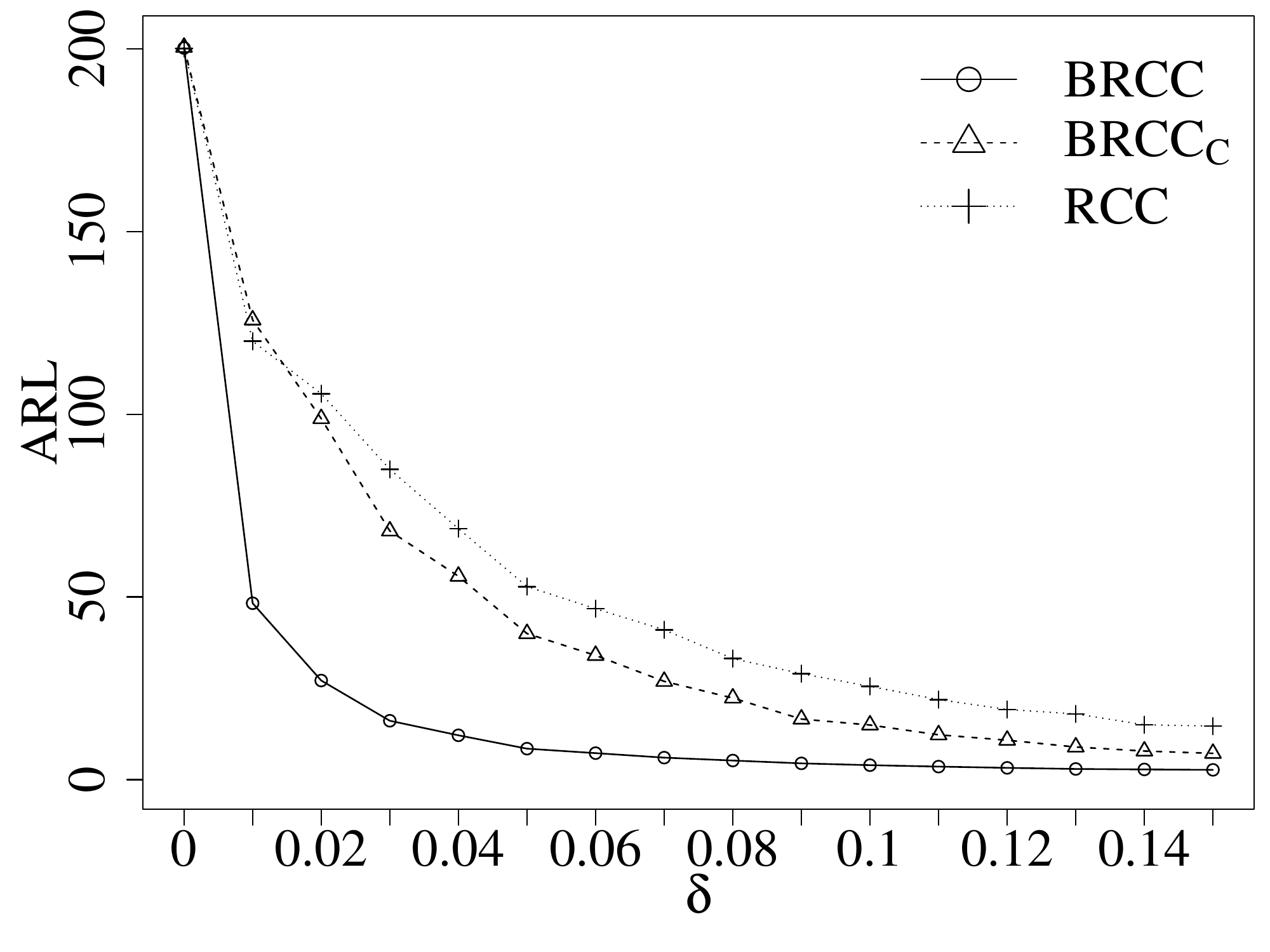}\label{f:6s}}  
\caption{ARL$_1$ when the dispersion is out-of-control, considering $n=200$. } 
\label{f:ARL200sig} 
\end{center} 
\end{figure} 

\begin{figure}
\begin{center} 
\subfigure[Scenario 1]{\includegraphics [width=8.0cm]{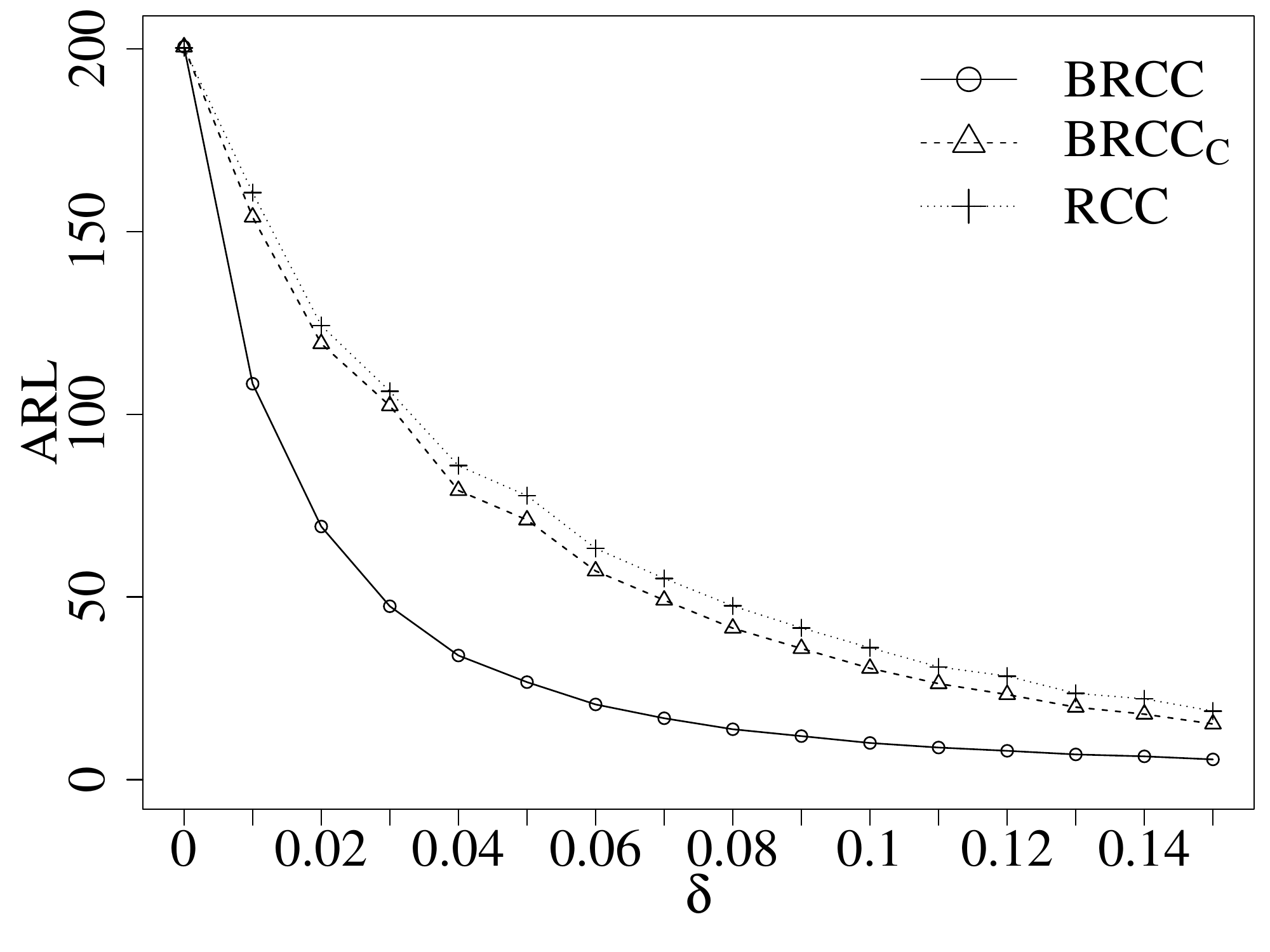}\label{f:1bs}} 
\subfigure[Scenario 2]{\includegraphics [width=8.0cm]{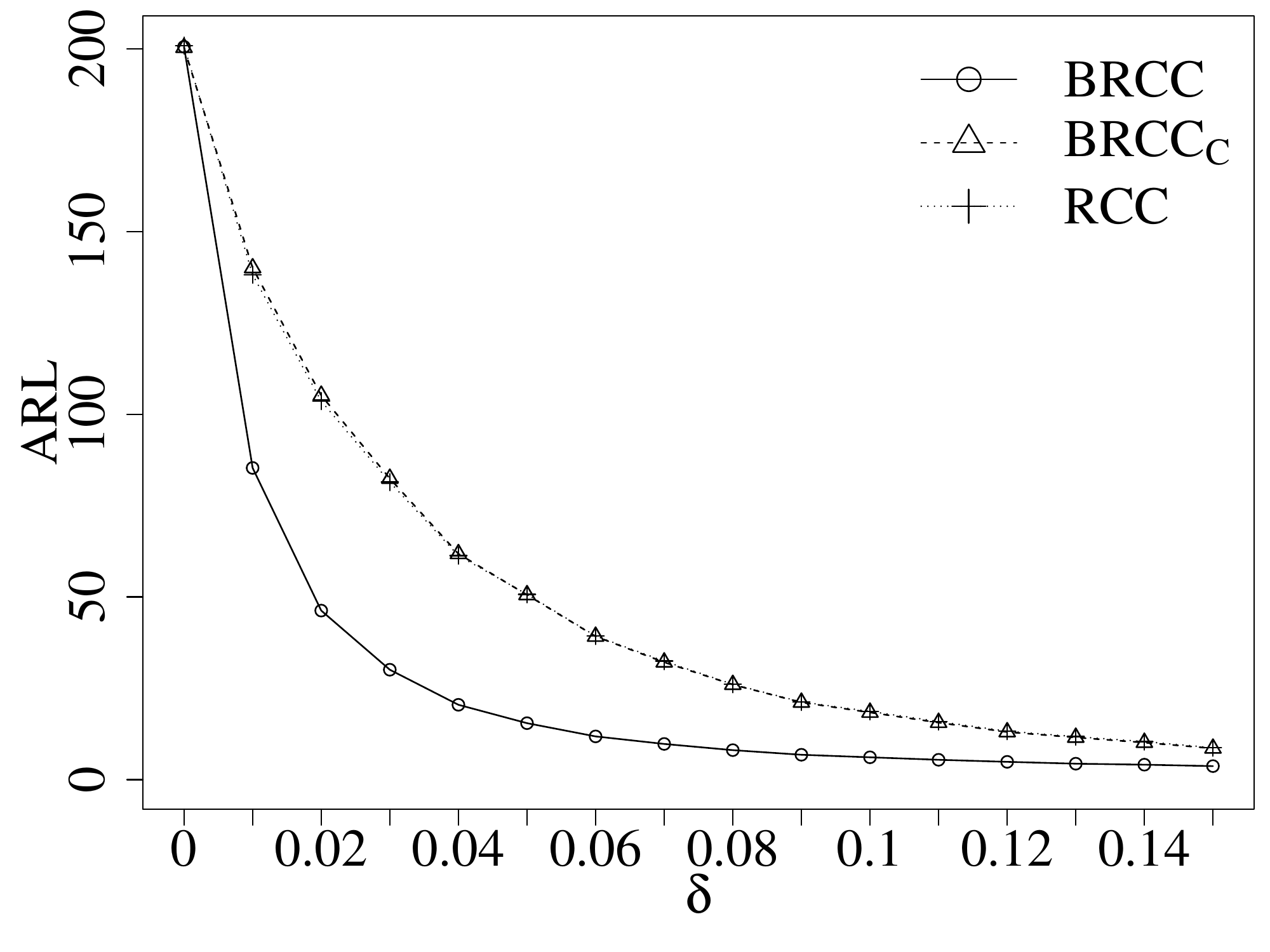}\label{f:2bs}} 
\subfigure[Scenario 3]{\includegraphics [width=8.0cm]{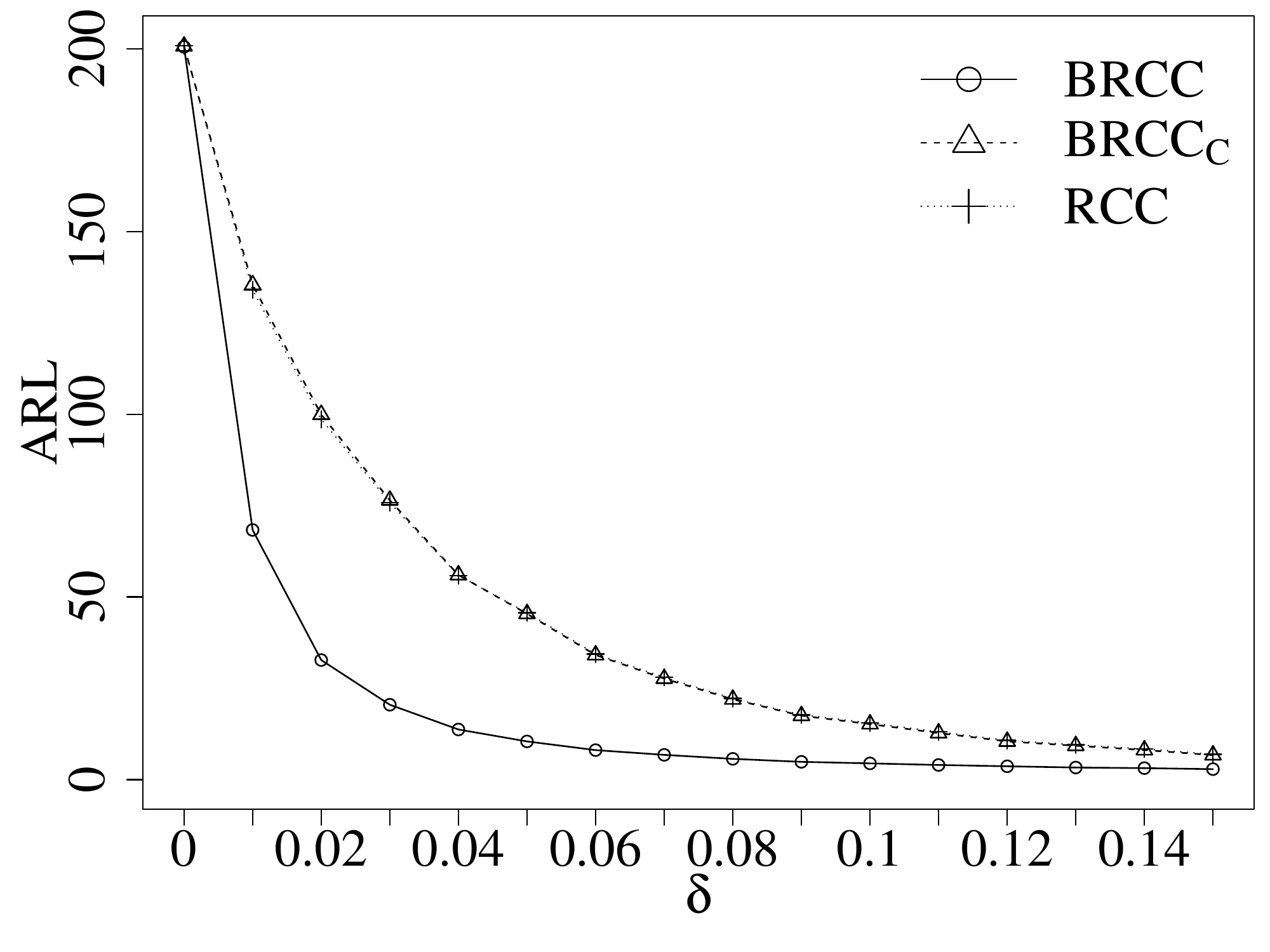}\label{f:3bs}} 
\subfigure[Scenario 4]{\includegraphics [width=8.0cm]{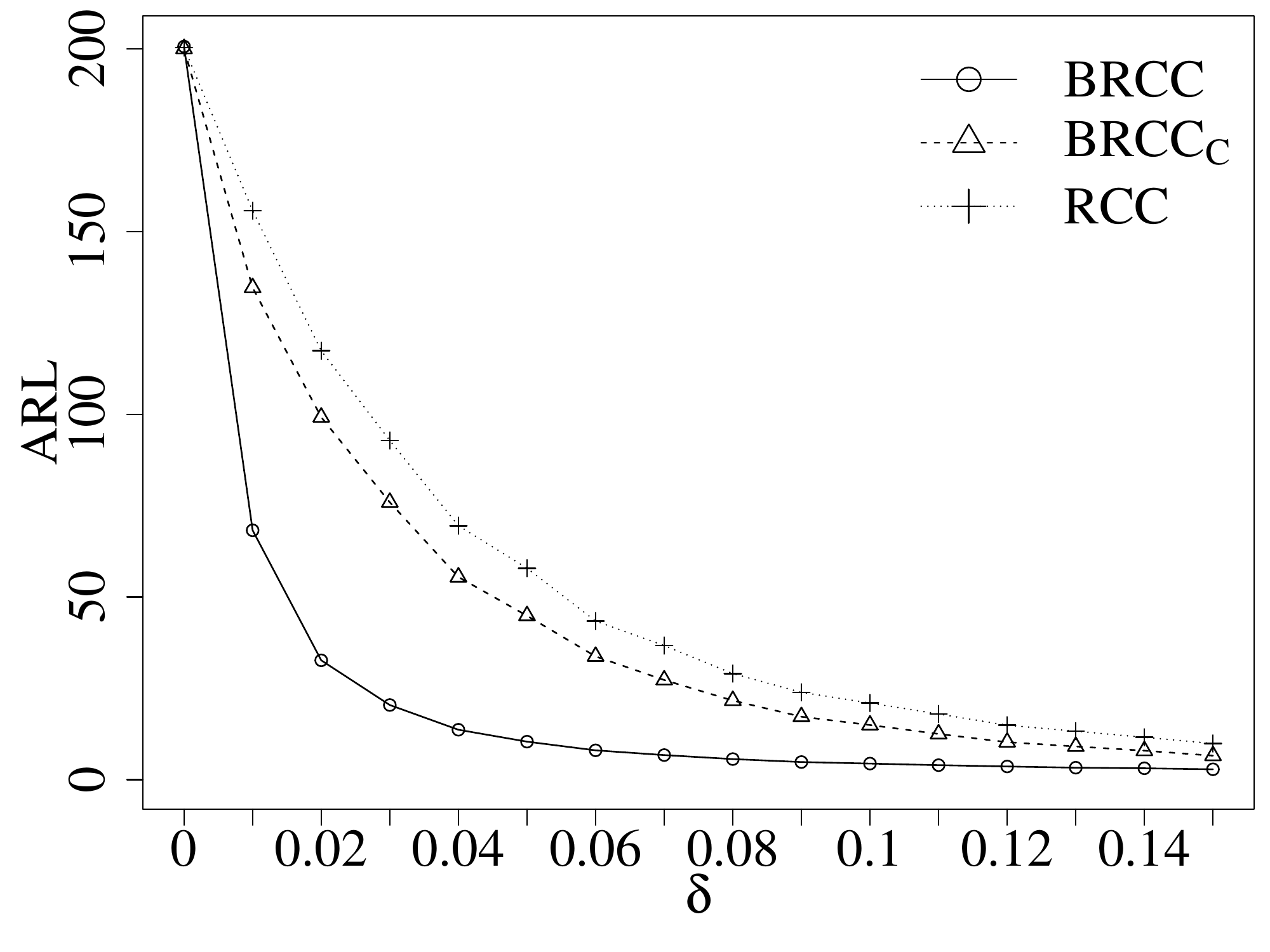}\label{f:4bs}} 
\subfigure[Scenario 5]{\includegraphics [width=8.0cm]{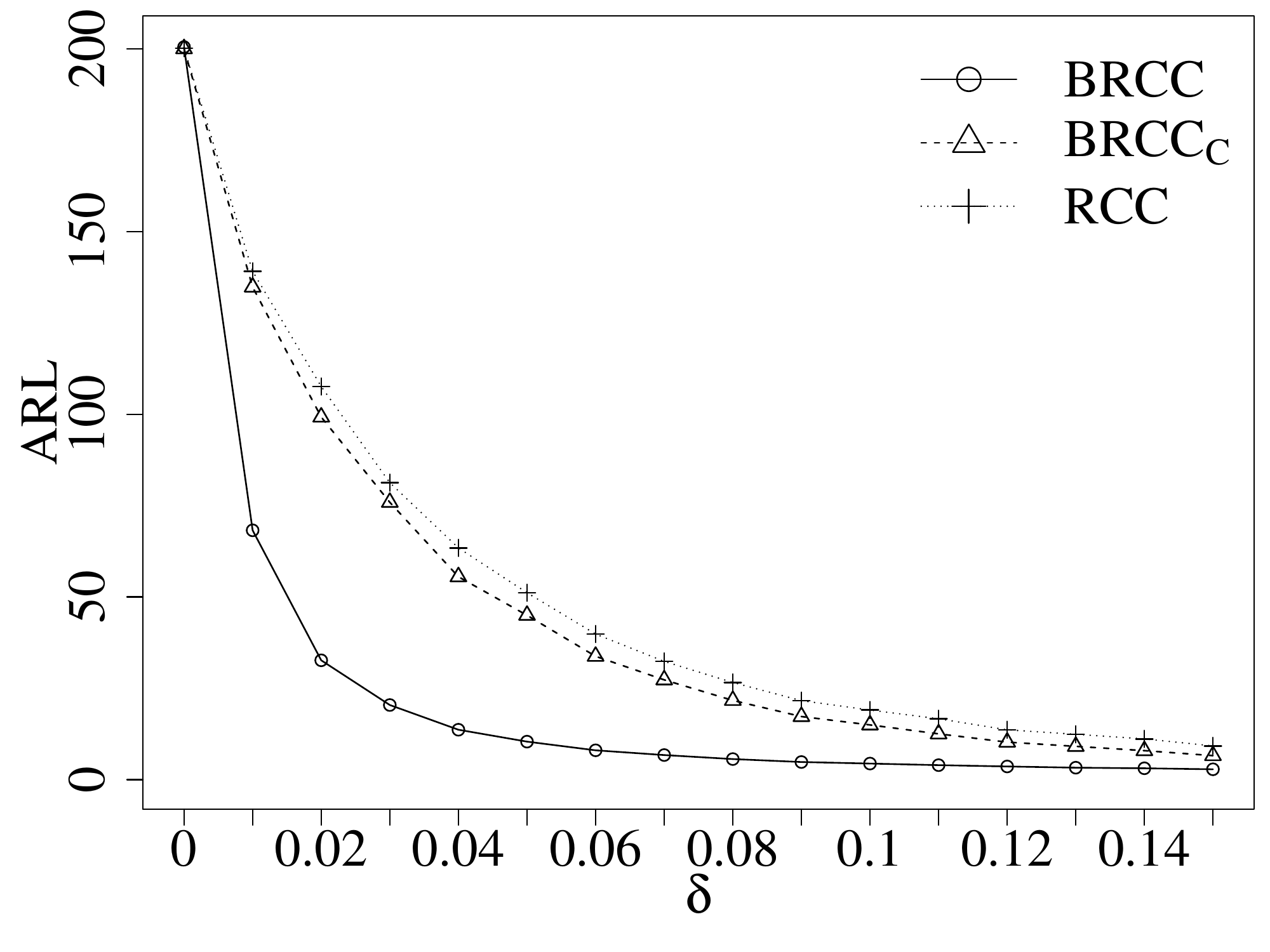}\label{f:5bs}}
\subfigure[Scenario 6]{\includegraphics [width=8.0cm]{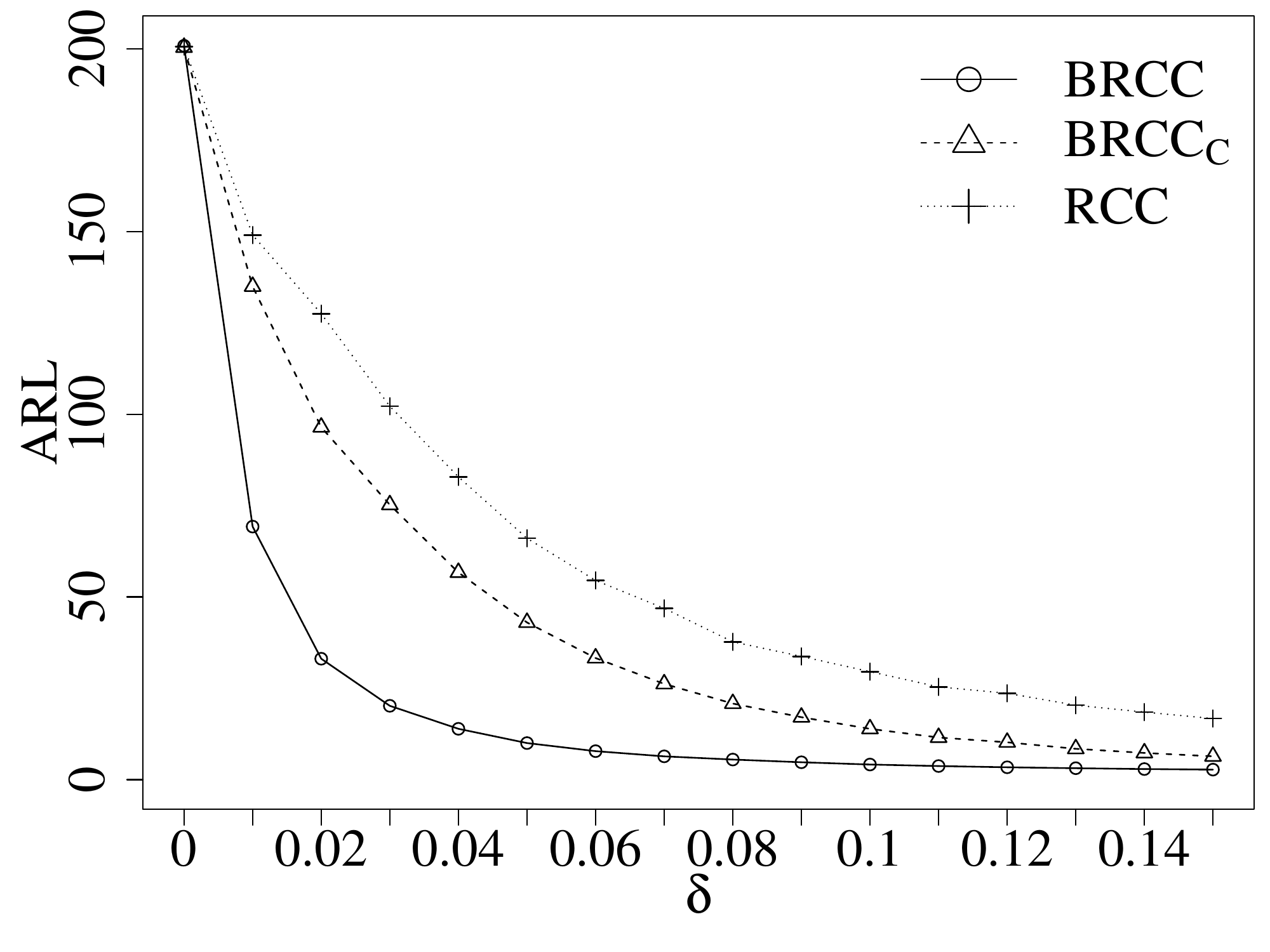}\label{f:6bs}}  
\caption{ARL$_1$ when the dispersion is out-of-control, considering $n=1000$.} \label{f:ARL1000sig} 
\end{center} 
\end{figure} 

An interesting result that favors the beta regression control charts is that when the process mean is closer to the extremes (Scenario $4$, Scenario $5$, and Scenario $6$), the RCC has values of ARL$_1$ higher than ARL$_0$. For $n$ = 200, these results are observed in Figures \ref{f:4}, \ref{f:5}, and \ref{f:6}; and for $n$ = 1000 in Figures \ref{f:4b}, \ref{f:5b}, and \ref{f:6b}. In Figure \ref{f:4b}, represented by a process $ \mu\cong 0.20$ and $ \sigma\cong 0.070$, the RCC presents an ARL$_1 \approx 237$ when $\delta = -0.02$, in Figure \ref{f:5b}, generated by a process of $ \mu\cong 0.80$ and $ \sigma\cong 0.070$, the value of ARL$_1 \approx 256$ when $ \delta= 0.4$, 
while in Figure \ref{f:6b}, generated by a process of $ \mu\cong 0.08$ and $ \sigma\cong 0.070$, the value of ARL$_1 \approx 587$ when $ \delta= -0.4$.  
These results reinforce the suggestion that the use of models where the data normality is assumed, such as linear regression, is inappropriate when the variable of interest is restricted in a limited range, such as rate and proportion variables, since the support of the normal distribution is the entire real line. Moreover, in situations similar to Scenarios $4$, $5$, and $6$, where the mean is at one of the ends, the RCC can result in limits outside the $0$ or $1$ interval, which contributes to distorted results in terms of ARL$_1$.

Figures \ref{f:ARL200sig} and \ref{f:ARL1000sig} show the results of ARL$_1$ evaluation when the dispersion is out-of-control. 
We can see that BRCC captures the dispersion increase more quickly than its competing charts, presenting the best performance in all considered scenarios. 
For the changes in dispersion the influence of the sample size in ARL$_1$ performance is negligible, as well as the results of changes in the mean of the process. 
 We highlight that 
the increase of variability of a process may imply an increase of defective units, whereas a reduction of the dispersion may indicate an increase in process capacity, since more units will be close to the correct specifications \citep{raey}.  

\begin{figure}[t] 
\begin{center} 
{\includegraphics [width=12cm]{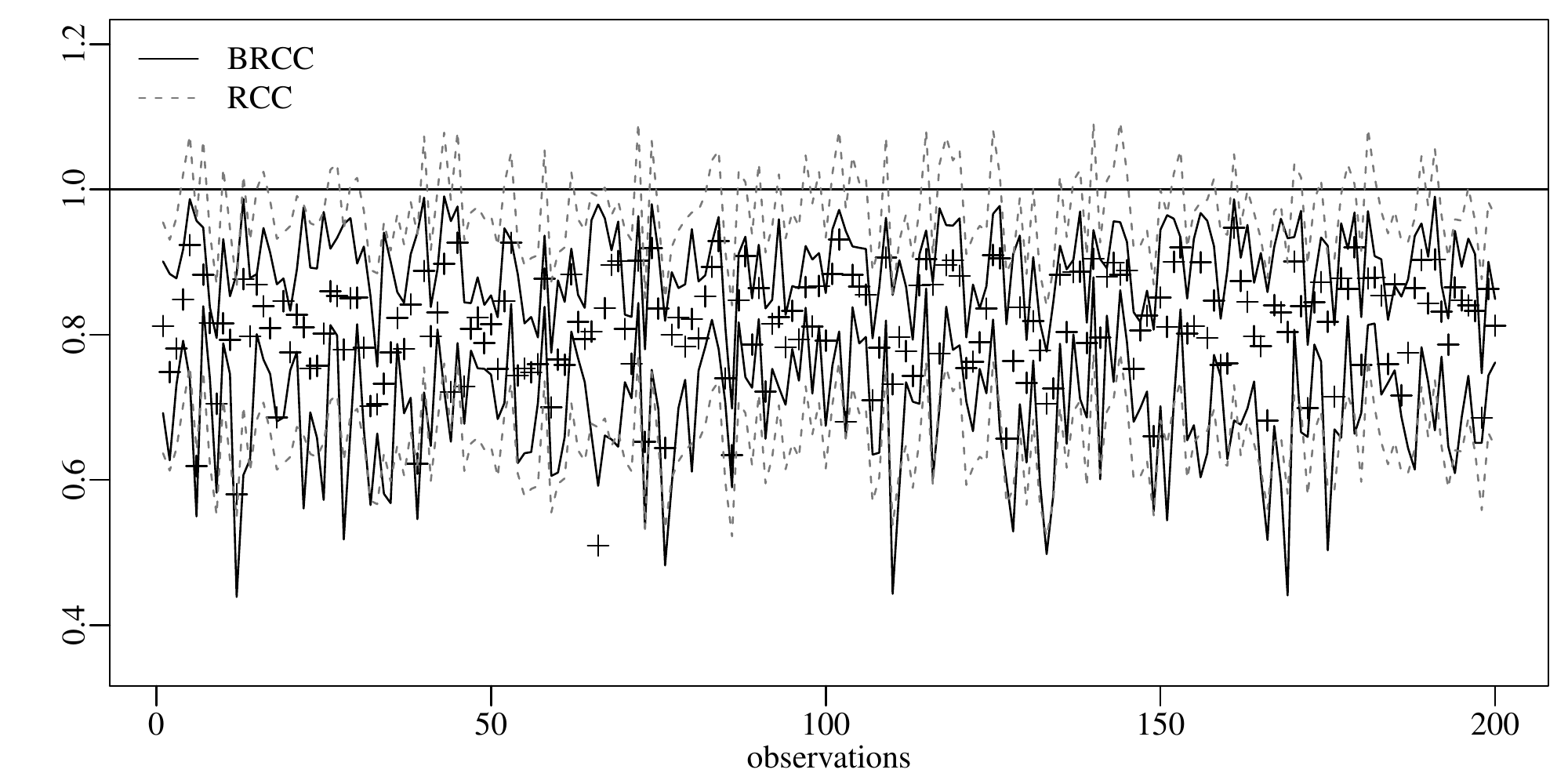}} 
\caption{RCC and BRCC for simulated data using the Scenario 5 ($\mu\cong 0.80$), with $n = 200$.} 
\label{simugrafic} 
\end{center} 
\end{figure}

Figure \ref{simugrafic} shows an application of the usual RCC and the proposed BRCC on a simulated data set. We note that the RCC has control limits above $1$, making no physical sense for fraction and proportion data type. In addition, we note that as the upper limit of the RCC exceeds the data upper limit, if the process goes out-of-control by increasing the average (getting closer to $1$) the RCC will tend to detect fewer out-of-control points. This experiment helps to explain the numerical results of ARL$_1$ in Figures  \ref{f:4}--\ref{f:5}--\ref{f:6} and \ref{f:4b}--\ref{f:5b}--\ref{f:6b}. Along with this, asymmetric generator processes increase the false alarm rate in RCC due to the discrepancy of the data with the normal distribution \citep{SantAnnaaandCaten:2012}, being that, rate and proportion data generally present asymmetry \citep{FerrariandCribari:2004}. It is possible to observe that the range of the BRCC limits is, in general, smaller than the range of the RCC limits, adapting itself better to the dispersion of each random variable throughout the observations. 

In general, the numerical results show the importance (i) of considering an adequate distribution for rate or proportion data and (ii) correctly modeling the dispersion structure in a regression control chart. In this way, quality engineers can jointly monitor the average and the dispersion of processes, properly detecting the occurrence of attributable causes. Thus, corrective actions can be taken before many nonconforming products are produced \citep{tsay}. 

\FloatBarrier

\section{Real data applications} \label{Exemplo} 

This section presents two empirical applications of the following control charts: BCC, RCC and BRCC. The first database is related to a tire manufacturing process and the second to relative air humidity data in the city of Bras\'ilia, Brazil. 

\subsection{Tire manufacturing process}\label{s:pneus} 

This application considers data referring to a radial tire manufacturing process of a multinational company of rubber products. This data is also considered in \cite{SantAnna:2009} and is available in Appendix A. The quality characteristic to be monitored ($y$) is the proportion of unconverted mass. This variable is obtained by the rate between the volume of raw material that was not converted into product and the total volume, being directly related to the loss of raw material. The control variables in this process are: tread length ($x_1$), section length ($x_2$), sidewall length ($x_3$), trim positioning ($x_4$) and kerf positioning ($x_5$). These five control variables are classified into three distinct classes, assuming values equal to $-1$, $0$ or $1$. 

To model this data, \cite{SantAnna:2009} fitted a beta regression model with constant dispersion considering the following covariates: $x_1$, $x_2$, $I_1$, $I_2$, and $I_3$, where $I_1$, $I_2$, and $I_3$ are interactions between the main variables defined by $I_1=x_1 \times x_2$, $I_2=x_1 \times x_4$, and $I_3=x_2 \times x_5$. 
The logit was the link function considered in both mean and dispersion submodels. 
In our application, the same control variables will be considered for the mean, both in the beta regression model and in the linear regression model. For the beta regression model with varying dispersion, after adjustments and tests, we consider the following control variables in the dispersion regression structure: $x_1$ and $I_1$. The adjusted model is presented in Table \ref{t:betareg-manufatura}. The calculated value of the likelihood ratio statistics (with $p$-value in brackets) for the constant dispersion test is ${\rm LR}=6.9016$ ($p\text {-value}=0.0317$). At the $5\%$ significance level, the test indicates that dispersion should be modeled. For this reason, the model with variable dispersion will be considered for the BRCC.

\begin{table}[t]
\caption{Beta regression model adjusted for tire manufacturing data.}
\label{t:betareg-manufatura} 
\begin{center}
\begin{tabular}{lrcrc}
\hline
& Estimate & Std. error & $z$ stat & $p\text{-value}$ \\
\hline
\multicolumn{5}{c}{Mean submodel}\\ 
\hline
Intercept & $	-3.5807	$ & $	0.2140	$ & $	-16.7300	$ & $	<0.0001	$ \\
$x_1	$ & $	0.4507	$ & $	0.2245	$ & $	2.0080	$ & $	0.0756	$ \\
$x_2	$ & $	0.4656	$ & $	0.2307	$ & $	2.0180	$ & $	0.0743	$ \\
$I_1	$ & $	-0.6716$ & $	0.2215	$ & $	-3.0340	$ & $	0.0142	$ \\
$I_3	$ & $	0.3054	$ & $	0.0185	$ & $	16.4990	$ & $	<0.0001	$ \\
$I_7	$ & $	0.2106	$ & $	0.0186	$ & $	11.3170	$ & $	<0.0001	$ \\
\hline
\multicolumn{5}{c}{Dispersion submodel}\\
\hline
Intercept & $	-3.0847	$ & $	0.2577	$ & $	-11.9690	$ & $	<0.0001	$ \\
$x_1	$ & $	-0.8563	$ & $	0.3659	$ & $	-2.3400	$ & $	0.0440	$ \\
$I_1	$ & $	0.8582	$ & $	0.3656	$ & $	2.3470	$ & $	0.0435	$ \\ 
\hline
\end{tabular} 
\end{center} 
\end{table}

\begin{figure}
\begin{center}
\includegraphics[width = 0.55\textwidth]{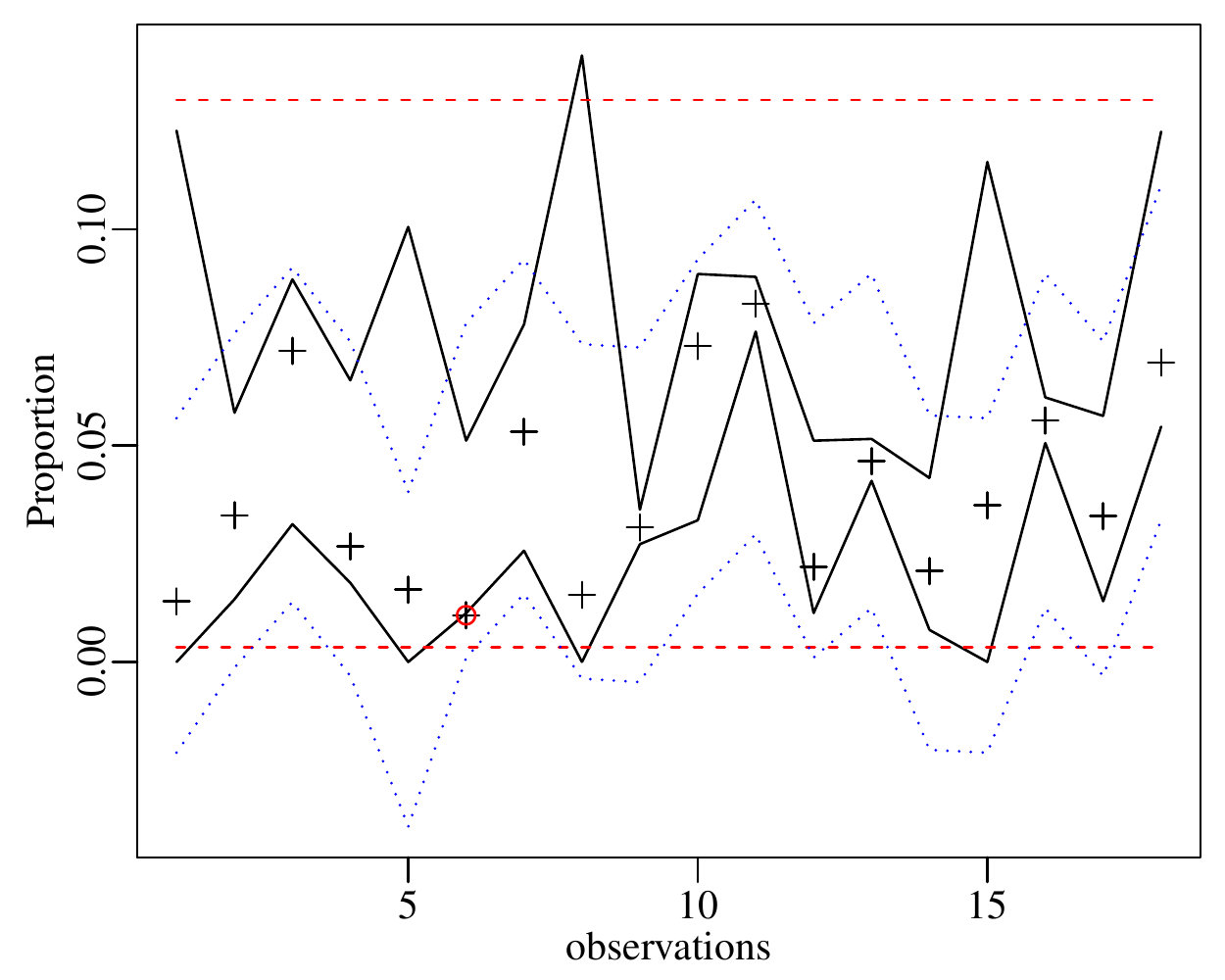} 
\caption{Control charts for tire manufacturing data. BCC in dashed red line, RCC in dotted blue line, and BRCC in solid black line.} 
\label{f:charts-manufatura} 
\end{center} 
\end{figure}

Figure \ref{f:charts-manufatura} presents the BCC, RCC and BRCC applied to the data of tire manufacturing. In order to build all control charts, it was fixed ARL$_0=200$, that is,  $\alpha=0.005$. It is noteworthy that BBC has control limits constant and fairly wide, not identifying out-of-control points. The usual RCC reduces the range of control intervals compared to BCC, but still considers all points as under control. It is noticed that some lower control limits defined by the RCC are below zero, making no practical sense and leading to loss of power in the detection of out-of-control points. On the other hand, the proposed BRCC, in general, presents intervals with smaller range than the two previous ones. By construction, due to the assumption of beta distribution to the control variable, the BRCC control limits are always within the $(0,1)$ range, being suitable for rate and proportion type variables. Considering the BRCC, we note that observation number six is out-of-control. When analyzing the characteristics of this observation, it is noted that the rate of unconverted mass is lower than expected for when the values of all covariates are equal to zero (according to the data in Appendix A). 

\FloatBarrier

\subsection{Air relative humidity} 

In this section 
an application of the proposed control chart for the relative humidity data (RH) of the city of Brasilia, capital of Brazil will be presented.  
The analyzed sampling period comprises from $06/21/2014$ to $06/19/2015$, making a total of $729$ daily observations. The variable of interest ($y$) is given by the rate of the water vapor amount contained in air and the maximum amount that could be contained at the same temperature, given the content was saturated \citep{Camuffo1998}. In this way, the RH value is presented as a percentage. 

The air relative humidity is an important variable in many areas of application such as environmental diagnosis and risk assessment \citep{Camuffo1998}. Low levels of RH can lead to fires, water stress, and health problems \citep{toxicology2002airliner}. High RH is also associated with several health problems, with respiratory, cardiac, and rheumatic symptoms. Studies also mention the relationship between fungi and mites with RH \citep{Arlian2001,Oreszczyn2006}. Meteorological variables, such as RH, increasingly affect and influence human health and may, together with other climatic factors, affect the incidence and distribution of infectious diseases. It can be seen that RH can indirectly affect the incidence and prevalence of allergic diseases and also plays a major role in the transmission of viral diseases, such as influenza \citep{Gao2014}, being an important variable for monitoring. 

Low levels of RH are observed in the city of Bras\'ilia during some periods of the year. Thus, monitoring this variable is important because preventive measures can be taken with regard to health care, water resources, and the agricultural sector. \textit{Dummy} variables will be used representing each season of the year, in order to monitor the RH average value in the city of Bras\'ilia. We chose to use these control variables because the region's larger differences in RH are observed during the different seasons of the year. The independent variable will assume $1$ in the presence of that season and $0$ otherwise. This procedure was performed for the summer, fall and winter seasons, with spring being considered the reference station. 

Table~\ref{t:betareg-RH} shows the adjusted beta regression model with varying dispersion for RH data. 
The logit was the link function considered in both mean and dispersion submodels. 
It is noticeable that all considered covariates are significant, both for the mean and dispersion submodels. The likelihood ratio test rejects the null hypothesis of constant dispersion because ${\rm LR} = 53.1020$ and $p$-value$ < 0.0001$. 
These same covariates were considered for the regression model in the RCC. The ARL$_0$ was fixed equal to $200$. 

\begin{table}[t]
\caption{Adjusted beta regression model with varying dispersion for Bras\'ilia RH data.}
\label{t:betareg-RH} 
\begin{center}
\begin{tabular}{lrcrc} 
\hline
& Estimate & Std. error & $z$ stat & $p\text{-value}$ \\
\hline
\multicolumn{5}{c}{Mean submodel}\\ \hline
Intercept & $	0.4255	$ & $ 0.0605 $ & $ 7.0370	$ & $ <0.0001 $\\ 
 Summer & $	0.5119	$ & $ 0.0781	$ & $	6.5580	$ & $ <0.0001 $\\	
 Fall & $ 0.2691 $ & $ 0.0767	$ & $	3.5070	$ & $ 0.0005 $\\ 
 Winter & $	-0.5073	$ & $ 0.0716 $ & $ -7.0820 $ & $ <0.0001 $\\ 
\hline
\multicolumn{5}{c}{Dispersion submodel}\\ \hline
Intercept & $-0.40319$ & $ 0.06817 $ & $ -5.915	$ & $ <0.0001 $\\
Summer & $	-0.45814 $ &$ 0.09516 $ & $	-4.814 $ & $ <0.0001 $\\
Fall & $ -0.43099$ & $ 0.09430 $ & $ -4.571	$ & $ <0.0001 $\\
Winter & $-0.62864 $ &$ 0.09283$ & $-6.772	$ & $ <0.0001 $\\ 
\hline
\end{tabular} 
\end{center} 
\end{table} 

\begin{figure}[h] 
\centering
\includegraphics[width=0.55\textwidth]{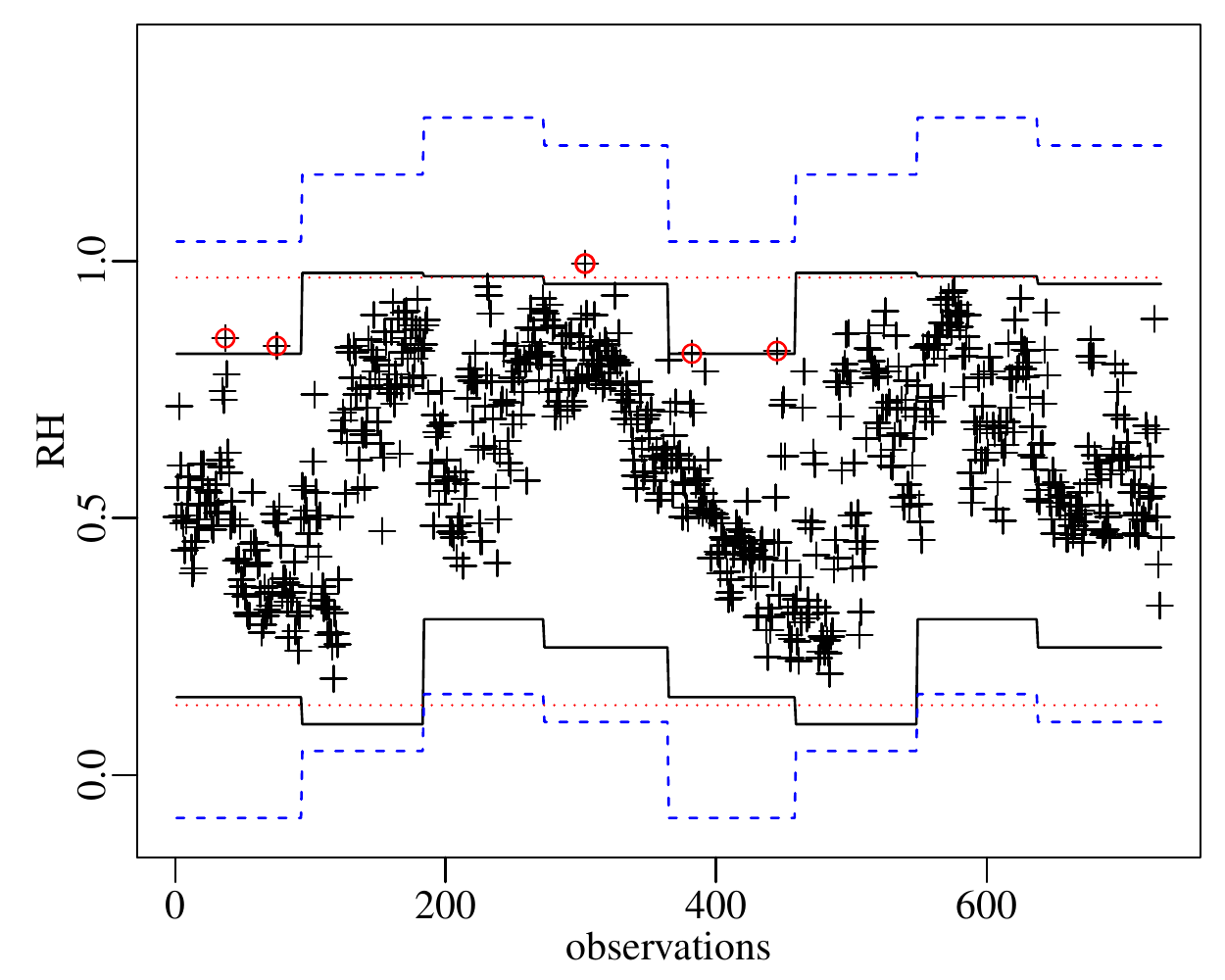}
\caption{Control chart for Bras\'ilia RH data. BCC in dashed red line, RCC in dotted blue line, and BRCC in solid black line.} 
\label{1fig}
\end{figure} 

The Figure \ref{1fig} shows the limits of BCC, RCC and BRCC to monitor the RH. As noted, the BCC points only to an out-of-control observation that corresponds to the fall season. Moreover, since they are constant limits, they do not follow the variability observed in the series. The limits of RCC are broader and therefore do not present any out-of-control observation. Another observed detail, which is corroborated by the previous numerical results, is that the limits are outside the $(0,1)$ range. In contrast, the BRCC showed narrower limits and greater sensitivity when identifying five atypical observations, four of which correspond to winter and one to fall. These observations present a very high RH value for the season in which they occurred. Bras\'ilia has two very different seasons: dry, from April to September; and rainy, from October to March. During the dry season, the air humidity can reach very low levels. On the other hand, during the rainy season the air humidity is much greater \citep{Menezzi2008}. The atypical observations correspond to the months of April, June and September that coincide with the dry season, where the RH should have low values and these observations range from $82\%$ to $99\%$. 

\FloatBarrier

\section{Conclusions}\label{S:conclusion} 

This work had the purpose of proposing a control chart useful for monitoring data of fraction or proportion type, restricted to the range $(0,1)$. The proposed control chart considers that the variable of interest has a beta distribution, in which the mean and dispersion parameters are modeled using regression structures involving control variables. In this way, the BRCC controls at the same time the mean and the dispersion of the variable of interest. 

There were two main approaches in literature to monitor double bounded data that is susceptible to control variables: i) The BCC, which assumes beta distribution (adequate for double bounded data) but, unlike the chart presented, does not consider quality characteristics that influence the variable of interest; and ii) the RCC, which considers quality characteristics that influence the process mean, but assumes the normality of the data, which is not an adequate assumption for variables of the fraction and proportion type. The proposed BRCC considers these two characteristics at the same time. 

To analyze the performance of the BRCC, an extensive simulation study was carried out, comparing it to BRCC$_C$ and RCC. The numerical results show the superiority of BRCC in terms of ARL, indicating a lower average number of samples until a real change in the process is detected. We also considered two applications to real data, considering BRCC, BCC, and RCC, evidencing the practical importance of the proposed method in the areas of quality control (tire manufacturing) and environmental science (relative humidity). 

The beta regression control chart with varying dispersion that was proposed is a promising technique for monitoring fraction or proportion type data in a production process. The proposed method is also useful in the monitoring of variables in different areas of application, such as hydrology, health, meteorology, among others.

\section*{Acknowledgements}

We gratefully acknowledge partial financial support from  
Conselho Nacional de Desenvolvimento Cient\'ifico e Tecnol\'ogico (CNPq), 
Coordena\c{c}\~ao de Aperfei\c{c}oamento de Pessoal de N\'ivel Superior (CAPES), 
and 
Funda\c{c}\~ao de Amparo a Pesquisa do Estado do Rio Grande do Sul (FAPERGS), Brazil. 
We thank the referees for their comments and suggestions.

\bibliographystyle{elsarticle-harv} 
\bibliography{Referenciais}

\newpage

\section*{Appendix A}

\begin{table}[h!]
\caption{Data from the tire manufacturing process presented in 
Section~\ref{s:pneus}.}
\centering
\begin{tabular}{cccccc}
  \hline
$y$ &  $x_1$ & $x_2$ & $x_3$ & $x_4$ & $x_5$ \\   
  \hline
  0.0140 &    -1 &    -1 &    -1 &    -1 &     1 \\ 
  0.0339 &    -1 &     1 &     1 &     1 &    -1 \\ 
  0.0719 &     1 &     1 &     1 &     1 &     1 \\ 
  0.0267 &     1 &     1 &    -1 &     1 &    -1 \\ 
  0.0167 &    -1 &    -1 &     1 &     1 &     1 \\ 
  0.0108 &     0 &     0 &     0 &     0 &     0 \\ 
  0.0532 &    -1 &     1 &    -1 &     1 &     1 \\ 
  0.0155 &    -1 &    -1 &     1 &    -1 &    -1 \\ 
  0.0311 &     1 &    -1 &     1 &    -1 &     1 \\ 
  0.0730 &    -1 &     1 &    -1 &    -1 &    -1 \\ 
  0.0828 &     1 &    -1 &     1 &     1 &    -1 \\ 
  0.0220 &     0 &     0 &     0 &     0 &     0 \\ 
  0.0464 &     1 &    -1 &    -1 &    -1 &    -1 \\ 
  0.0210 &     1 &     1 &     1 &    -1 &    -1 \\ 
  0.0362 &    -1 &    -1 &    -1 &     1 &    -1 \\ 
  0.0558 &     1 &    -1 &    -1 &     1 &     1 \\ 
  0.0337 &     1 &     1 &    -1 &    -1 &     1 \\ 
  0.0692 &    -1 &     1 &     1 &    -1 &     1 \\
   \hline
\end{tabular}
\end{table}

\end{document}